\documentclass{article} %
\usepackage{iclr2027_conference,times}

\usepackage{amsmath,amsfonts,bm}

\def\eqref#1{equation~\ref{#1}}

\def\1{\bm{1}}

\DeclareMathAlphabet{\mathsfit}{\encodingdefault}{\sfdefault}{m}{sl}
\SetMathAlphabet{\mathsfit}{bold}{\encodingdefault}{\sfdefault}{bx}{n}

\usepackage{hyperref}
\usepackage{url}
\usepackage{enumitem}
\usepackage{booktabs}
\usepackage{graphicx}
\usepackage{soul}
\usepackage{wrapfig}
\usepackage{multirow}

\title{DiffStent: Surrogate Adjoint Differentiation for Stent Deployment in Anisotropic Arteries}

\author{Shoaib A. Goraya\textsuperscript{1,$\ast$}, Ajay Manicka\textsuperscript{1,$\ast$}, Aviral Jain\textsuperscript{2,$\ast$}, \\
Karim Kadry$^{1}$, Samuel Ruip\'erez-Campillo$^{1}$, Elazer R. Edelman$^{1}$ \& Farhad R. Nezami$^{2}$\\
$^{1}$Massachusetts Institute of Technology, Cambridge, MA 02139, USA \\
$^{2}$Brigham and Women's Hospital, Harvard Medical School, Boston, MA 02115, USA \\
\texttt{\{goraya, ajay\_m, kkadry, sruipere, ere\}@mit.edu} \\
\texttt{\{ajain30, frikhtegarnezami\}@bwh.harvard.edu}
}

\iclrfinalcopy %
\begin{document}
\raggedbottom %

\maketitle
\lhead{Preprint. Under review.}
\thispagestyle{fancy}
{\renewcommand{\thefootnote}{$\ast$}\footnotetext{Equal contribution.}}

\begin{abstract}
We present \textbf{DiffStent}, a differentiable finite element framework for simulating arterial deformation during in silico stent deployment, where nearly incompressible, anisotropic vessel walls undergo large deformation and frictional contact with the stent. DiffStent combines the fiber-reinforced Holzapfel–Gasser–Ogden constitutive model with an implicit discrete adjoint that differentiates the converged mechanical equilibrium without unrolling the nonlinear solver, thus reducing the memory cost by three orders of magnitude. We identify three mechanisms that can lead to inaccurate, unstable, or uninformative gradients: nonsmooth switching, state mismatch, and gradient cancellation. To address these, we apply a contact surrogate in the backward pass while retaining the original physical model in the forward pass. We show superior gradient accuracy of our method over AutoDiff and finite differencing across constitutive benchmarks, idealized stented vessels, and patient-specific calcified coronary anatomies, and use the resulting sensitivities to quantify how tissue properties and calcification patterns govern lumen expansion and arterial wall stress. Finally, we demonstrate that DiffStent enables gradient-based system identification from retrospective clinical images of atherosclerotic coronary anatomy. Together, these results establish a practical approach for differentiating through nonlinear biomechanical simulations with stiff constitutive laws and nonsmooth contact, enabling patient-specific parameter inference and optimization in computational cardiovascular mechanics.

\end{abstract}

\section{Introduction}

Predicting how soft tissues deform under mechanical contact is necessary for planning interventions, designing endovascular devices, and simulating surgery. These deformations depend on the tissue's material properties and fiber architecture: arteries distend under pressure \citep{taheri2026structure}, ventricular myocardium twists during a cardiac pulse \citep{li2025effects}, and skeletal muscle bulges during contraction \citep{guzman2025mri}. The properties governing these responses are difficult to measure directly, but can be inferred from observed deformation. This inference requires determining how sensitive a simulated result, like lumen gain, is to changes in tissue properties. One method of obtaining these sensitivities is differential simulation, which enables efficient gradient-based optimization to recover material information from a forward model deployment. Although differentiable simulators have been developed for rigid bodies, articulated robots, fluids, and elastic solids \citep{hu2019difftaichi, Qiao2020Scalable, du2021diffpd, huang2021plasticinelab, li2022diffcloth, li2024neuralfluid}, computing reliable sensitivities for anisotropic tissues undergoing large deformation and device contact remains challenging.

For anisotropic tissues, these sensitivities must account for cross-linking fiber orientations, spatially varying material properties, and mechanical loads transmitted through device contact. Existing methods address parts of this problem. Neural Modular Physics \citep{li2025neural} learns constitutive and time-integration modules and adapts to unknown material behavior and external forces from observed motion, but does not provide accurate gradients for fiber-reinforced anisotropic tissues interacting with devices. Fiber-reinforced constitutive laws \citep{holzapfel2000new, gasser2005hyperelastic, holzapfel2009constitutive} govern tissue mechanics, but merely computing deformation does not provide the sensitivities needed to infer tissue parameters from observed deformation or to select a balloon pressure that achieves the desired lumen expansion while limiting wall stress. Unrolling every step of a physics solver to obtain these gradients is computationally prohibitive as the cost  scales with the number of model parameters. Moreover, exponential stiffening and mathematical discontinuities can compromise simulation gradients \citep{suh2022differentiable}. Recent contact methods respond with surrogate gradients that sidestep the pathologies of the forward model \citep{paulus2026differentiable}. However, these models face additional difficulties within the tissue: tension-only fiber recruitment can make the tangent stiffness discontinuous, while near-incompressibility can make the equilibrium system ill-conditioned. During stent deployment, these material effects interact with contact activation as the balloon expands the stent against the artery. We therefore ask: \emph{can we develop a method for accurate yet efficient computation of gradients in fiber-reinforced tissues under contact to enable tissue parameter identification from observed deformation and balloon-pressure optimization for a desired clinical outcome?}

\begin{figure*}[t]
    \centering
    \includegraphics[width=\textwidth]{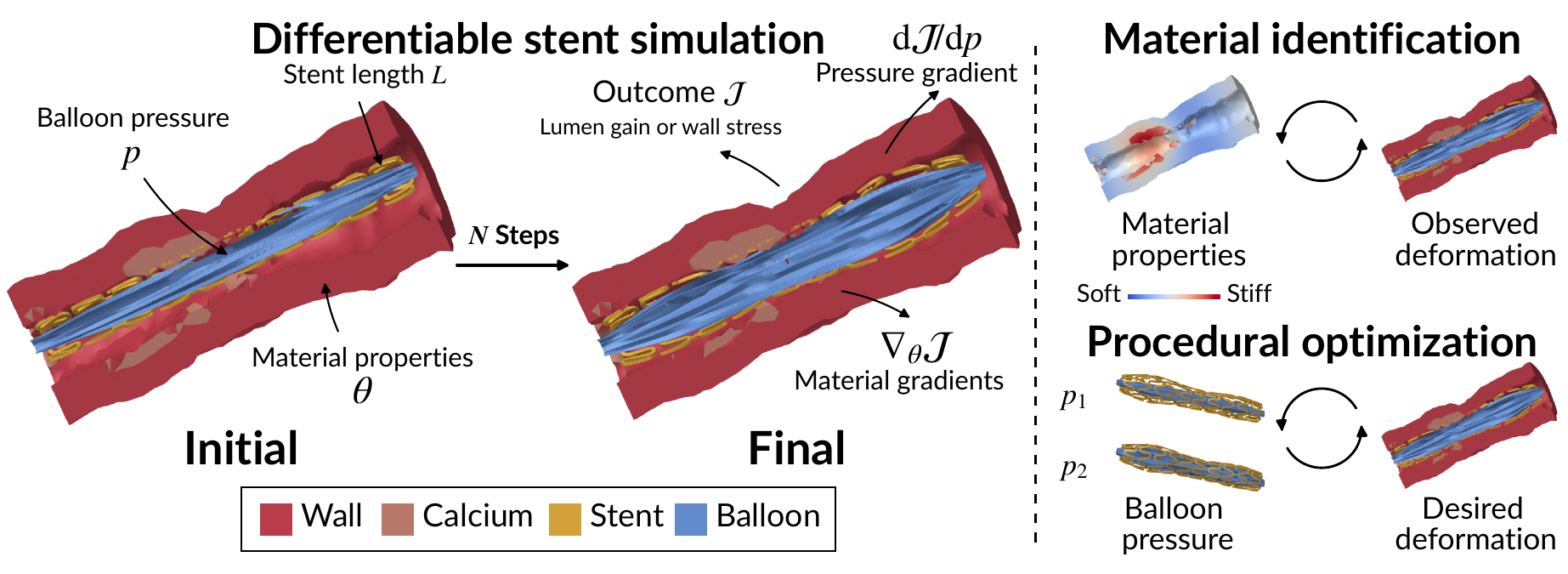}
    \caption{\textbf{DiffStent supports stent deployment simulation, material property estimation, and procedural optimization.} The simulator couples balloon, stent, and arterial mechanics in patient-specific vessels and computes outcome sensitivities w.r.t tissue properties and balloon pressure. These gradients support material estimation from clinically-observed deformation and contact pressure optimization for desired lumen expansion subject to stress constraints.}
    \label{fig:overview}
\end{figure*}

We present \textbf{DiffStent} (Figure \ref{fig:overview}), a GPU-native differentiable finite element solver for multiple deformable bodies in frictional contact. During deployment, the inflating balloon loads the stent, which expands against the artery, coupling all three bodies through balloon--stent and stent--artery contact. The multilayer arterial wall follows the nearly incompressible, anisotropic Holzapfel--Gasser--Ogden (HGO) constitutive law with fiber dispersion \citep{holzapfel2000new, gasser2005hyperelastic}, which represents its collagen families and reproduces its exponential stiffening under stretch. The forward simulations are validated against stent-expansion benchmarks (Appendix~\ref{app:forward-validation}).

For efficient gradient computation, we implement an adjoint formulation based on the implicit function theorem at the converged equilibrium state, significantly reducing memory cost by making gradient computation independent of the number of relaxation steps. We show superior gradient accuracy of this method over finite differencing and AutoDiff by applying the backward-pass surrogate to the three pathologies that corrupt gradient computations in this class of problems: \textit{nonsmooth switching} increases the tangent stiffness as fibers recruit or contact activates; \textit{state mismatch} arises when recorded states retain residual forces or the adjoint lacks solution history dependence; \textit{gradient cancellation} amplifies numerical error when derivative contributions nearly cancel. The proposed backward pass improves sensitivities in all three pathologies while preserving the forward mechanics (Section~\ref{sec:theory-pathologies}). 

Our contributions are as follows:

\begin{itemize}[leftmargin=1.2em, itemsep=1.5pt, topsep=3pt, parsep=0pt]
  \item A differentiable finite element simulator for multibody frictional contact with a nearly incompressible, anisotropic tissue model.
  \item A discrete adjoint formulation with a backward-pass contact surrogate and an analysis of gradient reliability under nonsmooth switching, state mismatch, and gradient cancellation.
  \item Gradient-based method for tissue parameter identification and deployment-protocol optimization on an idealized and patient-specific coronary artery.
\end{itemize}

\section{Background}
\label{sec:background}

\noindent\textbf{Differentiable and Learned Simulation.}
Differentiable simulation enables gradient-based identification, design, and control~\citep{newbury2024review}, with applications ranging from tactile material inference to joint geometry and control optimization in fluidic systems~\citep{si2024difftactile, li2024neuralfluid}. These tasks require sensitivities, computed either by unrolling solver iterations or by implicitly differentiating the equations defining their solution~\citep{bolte2023one, wang2025dilqr}. Progressively refining linear solves can reduce the cost of either strategy~\citep{bhatia2025prdp}. Complementary neural approaches learn physical states or evolution operators for prediction on general geometries and scalable rigid-body simulation~\citep{wu2024transolver, rubanova2024learning}. Physics-informed representations go further by incorporating mechanics through anisotropic energy minimization over a shell or by learning constitutive and integration modules for elastic dynamics~\citep{kairanda2024neuralclothsim, li2025neural}. Such representations support state prediction or equilibrium computation, but neither task alone establishes accurate material sensitivities. We compute these sensitivities through the converged finite element equilibrium of the prescribed anisotropic HGO material model, using an implicit adjoint to circumvent backpropagating through dynamic relaxation.

\noindent\textbf{Gradients through Nonsmooth and Stiff Physics.}
Difficulties in gradient computation arise from both nonsmooth constitutive laws and numerical errors on smooth branches. For nonsmooth contractive fixed-point problems, generalized derivatives support convergence analyses of iterative and implicit differentiation under suitable regularity assumptions~\citep{grazzi2024nonsmooth}. In contact simulation, adaptive differentiation horizons and selective substitution of value-function gradients mitigate errors around stiff events and nonsmooth transitions~\citep{georgiev2024adaptive, gao2024adaptive}. Other methods supply gradients before contact through long-range potentials~\citep{ye2026efficient} or backward-only straight-through contact models, with adaptive time integration improving hard-contact derivative accuracy~\citep{paulus2026differentiable}. Even on smooth branches, approximate forward and adjoint solves introduce distinct gradient errors~\citep{bhatia2025prdp}. Conditioning also matters: ill-conditioned differential operators can impair physics-informed training~\citep{rathore2024challenges}, while learned preconditioners improve discretized PDE solves~\citep{yang2025learning}. In stent deployment, fiber recruitment and contact introduce switches, while residual forces and omitted history can make equilibrium sensitivities differ from those of the simulated response. Cancellation between derivative contributions can further amplify numerical error. We analyze these mechanisms in Section~\ref{sec:theory-pathologies}, distinguishing the role of a backward-pass contact surrogate from the remaining limits on gradient reliability.

\begin{figure*}[htbp]
    \centering
    \includegraphics[width=\textwidth]{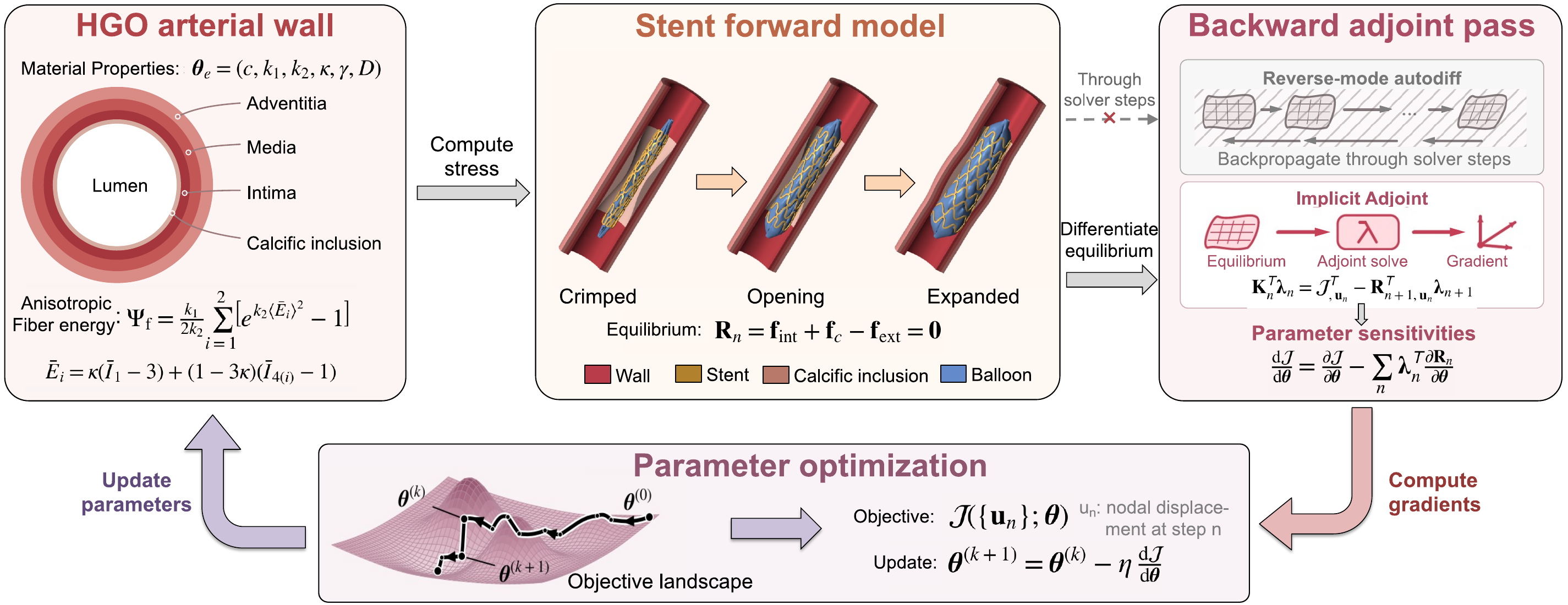}
    \caption{\textbf{DiffStent computational framework.} Geometry, material properties, and loading define a coupled balloon--stent--artery simulation. An equilibrium adjoint computes sensitivities of scalar outcomes, with a contact surrogate applied only during gradient computation. These sensitivities support analysis of tissue properties and evaluation of gradient-based deployment optimization. In the figure, comma subscripts denote partial derivatives, and the superscript ${}^{\top}$ is transpose of the entire derivative, as written explicitly in~\eqref{eq:adjoint}.}
    \label{fig:framework}
\end{figure*}

\section{Method}
\label{sec:theory}

DiffStent maps tissue properties and mechanical balloon loading to the coupled expansion of the stent and deformation of the arterial wall. We compute how changes in these inputs affect scalar outcomes such as lumen area or mean wall stress (Section~\ref{sec:theory-parameters}). We formulate this map through conservation of momentum (Section~\ref{sec:theory-discrete}) and compute solution sensitivities with an implicit adjoint pass (Section~\ref{sec:theory-adjoint}), avoiding backpropagation through the forward solver's relaxation steps. Targeted backward-pass corrections provide surrogate gradients near contact activation while preserving the forward mechanics (Section~\ref{sec:theory-pathologies}). Our implementation uses NVIDIA Warp~\citep{warp2022} and is visualized in Figure~\ref{fig:framework}.
 
\subsection{Parameters and objectives}
\label{sec:theory-parameters}
Let $\boldsymbol{\theta}$ represent the tissue parameters and $\boldsymbol{\beta}=(\beta_1,\ldots,\beta_N)$ represent the balloon pressures at $N$ recorded load steps. The simulator produces nodal displacements $\mathbf{u}_n$ at each load step $n$. The scalar functional $\mathcal{J}(\{\mathbf{u}_n\};\boldsymbol{\theta},\boldsymbol{\beta})$ represents the outcome of interest. We compute its derivatives with respect to selected material or loading parameters. Material sensitivities describe how assumed tissue properties affect deployment. Pressure sensitivities enter the constrained optimization of lumen area subject to stress limits in, and these gradients also provide the mathematical framework for material identification and procedural optimization (Fig. ~\ref{fig:identification_optimization}).

\subsection{Differentiable equilibrium model}
\label{sec:theory-constitutive}
\label{sec:theory-discrete}

The arterial wall follows the Holzapfel--Gasser--Ogden (HGO) anisotropic hyperelastic material model~\citep{holzapfel2000new,gasser2005hyperelastic}. At a reference position $\mathbf{X}$, the displacement field $\mathbf{u}$ defines the deformation gradient $\mathbf{F}=\mathbf{I}+\nabla_{\mathbf{X}}\mathbf{u}$, where $\mathbf{I}$ is the identity tensor. (Figure~\ref{fig:nonlinear-mapping} illustrates the mapping between configurations). The determinant $J=\det\mathbf{F}>0$ measures the local ratio of deformed to reference volume. The right Cauchy--Green tensor $\mathbf{C}=\mathbf{F}^{\top}\mathbf{F}$ describes local stretch, while its isochoric counterpart $\bar{\mathbf{C}}=J^{-2/3}\mathbf{C}$ removes uniform volume change to isolate shape distortion. The invariants $\bar I_1=\operatorname{tr}\bar{\mathbf{C}}$ and $\bar I_{4(i)}=\mathbf{a}_{0i}^{\top}\bar{\mathbf{C}}\mathbf{a}_{0i}$ measure matrix and fiber deformation, respectively. Here, $\mathbf{a}_{0i}$, $i=1,2$, are unit vectors along the mean orientations of the two collagen fiber families. The families lie in the local circumferential--axial plane at angles $\pm\gamma$ relative to the circumferential direction. Their strain measures and the energy per unit reference volume are
\begin{align}
\bar E_i &= \kappa(\bar I_1-3)+(1-3\kappa)(\bar I_{4(i)}-1),
\label{eq:fiberstrain}\\
\Psi(\mathbf{C};\boldsymbol{\theta})
&= \underbrace{\vphantom{\frac{k_1}{2k_2}\sum_{i=1}^{2}\left[\exp\!\left(k_2\langle\bar E_i\rangle^2\right)-1\right]}\frac{1}{D}\left(\frac{J^2-1}{2}-\ln J\right)}_{\Psi_{\mathrm{vol}}}
+\underbrace{\vphantom{\frac{k_1}{2k_2}\sum_{i=1}^{2}\left[\exp\!\left(k_2\langle\bar E_i\rangle^2\right)-1\right]}\frac{c}{2}(\bar I_1-3)}_{\Psi_{\mathrm{gm}}}
+\underbrace{\frac{k_1}{2k_2}\sum_{i=1}^{2}\left[\exp\!\left(k_2\langle\bar E_i\rangle^2\right)-1\right]}_{\Psi_{\mathrm{f}}}.
\label{eq:energy}
\end{align}
The volumetric energy $\Psi_{\mathrm{vol}}$ penalizes volume change, while $\Psi_{\mathrm{gm}}$ and $\Psi_{\mathrm{f}}$ describe the ground-matrix and fiber contributions, respectively. These latter terms depend on the isochoric deformation and together generate the deviatoric part of the Cauchy stress. Here $c$ and $k_1$ are matrix and fiber stiffnesses, $k_2$ controls exponential stiffening, and $D$ sets the bulk modulus $K=2/D$. The dispersion parameter $\kappa\in[0,1/3]$ ranges from aligned ($\kappa=0$) to isotropically distributed fibers ($\kappa=1/3$). The Macaulay bracket $\langle x\rangle=\max(x,0)$ restricts fiber loading to tension. The parameters $\boldsymbol{\theta}_e=(c,k_1,k_2,\kappa,\gamma,D)$ are assigned by layer for the idealized vessel or element $e$ for the patient-specific coronary segment (Section~\ref{sec:exp-setup}).

Finite element discretization couples the wall, stent, and balloon through penalty contact and regularized friction. At a recorded load step, the quasi-static equilibrium satisfies
\begin{equation}
\mathbf{R}_n(\mathbf{u}_n;\boldsymbol{\theta},\beta_n,\mathbf{h}_n)
=\mathbf{f}_{\mathrm{int}}(\mathbf{u}_n;\boldsymbol{\theta},\mathbf{h}_n)+\mathbf{f}_{c}(\mathbf{u}_n)-\mathbf{f}_{\mathrm{ext}}(\beta_n)=\mathbf{0},
\label{eq:residual}
\end{equation}
where $\mathbf{f}_{\mathrm{int}}$, $\mathbf{f}_{c}$, and $\mathbf{f}_{\mathrm{ext}}$ are internal, contact, and applied forces. The internal force depends on the displacement $\mathbf{u}_n$ and plastic-history state $\mathbf{h}_n$ which carries plastic strain and hardening information from prior loading in the stent and yielding plaque. This state evolves during deployment and affects the stress at a given displacement. The external force $\mathbf{f}_{\mathrm{ext}}(\beta_n)$ represents the force applied from the balloon to the stent. The forward solver approaches equilibrium by explicit dynamic relaxation, and residual force measures the accuracy of each recorded state. Contact and device models, constitutive derivations, and numerical settings are given in Appendices~\ref{app:formulation}, \ref{app:derivations}, and~\ref{app:hyperparams}, respectively.

\subsection{Discrete adjoint sensitivities}
\label{sec:theory-adjoint}

At equilibrium, the residual $\mathbf{R}_n=\mathbf{0}$ defines the displacement $\mathbf{u}_n$ implicitly in terms of the tissue parameters $\boldsymbol{\theta}$. For the chained derivative, we substitute the plastic-history updates into the residual and objective; their derivatives then include the resulting dependence on the displacement history. Let $\mathbf{K}_n=\partial\mathbf{R}_n/\partial\mathbf{u}_n$ denote the corresponding Jacobian. For a fixed initial state, differentiable updates, and nonsingular Jacobians, we solve for the adjoint variables $\boldsymbol{\lambda}_n$ backward over load steps $n=N,\ldots,1$, with terminal condition $\boldsymbol{\lambda}_{N+1}=\mathbf{0}$, and assemble the parameter gradient:
\begin{equation}
\mathbf{K}_n^{\top}\boldsymbol{\lambda}_n
=\left(\frac{\partial\mathcal{J}}{\partial\mathbf{u}_n}\right)^{\!\top}
-\sum_{m=n+1}^{N}\left(\frac{\partial\mathbf{R}_{m}}{\partial\mathbf{u}_n}\right)^{\!\top}\boldsymbol{\lambda}_{m},
\qquad
\frac{\mathrm{d}\mathcal{J}}{\mathrm{d}\boldsymbol{\theta}}
=\frac{\partial\mathcal{J}}{\partial\boldsymbol{\theta}}
-\sum_{n=1}^{N}\boldsymbol{\lambda}_n^{\top}\frac{\partial\mathbf{R}_n}{\partial\boldsymbol{\theta}}.
\label{eq:adjoint}
\end{equation}
The sum is empty at $n=N$ and reduces to the next-step term shown in Figure~\ref{fig:framework} when only successive states are coupled. The endpoint adjoint holds recorded history fixed, so the cross-step terms vanish. Pressure sensitivities follow by replacing parameter derivatives with derivatives with respect to $\boldsymbol{\beta}$, holding $\boldsymbol{\theta}$ fixed. Section~\ref{sec:exp-ablations} compares endpoint and chained sensitivities.

For each scalar objective (lumen gain, contact pressure etc.), the adjoint requires one transposed solve per recorded step, independent of the number of parameters. It stores recorded states rather than the full relaxation trajectory. Automatic differentiation (AD) of the recorded residual supplies the endpoint tangent, recovered by sparse probing and factorized in double precision (Appendix~\ref{app:adjoint}). These equilibrium derivatives require adequately relaxed states. Residual forces and omitted history dependencies can limit agreement with derivatives of the simulation, particularly during deflation.

\subsection{Gradient pathologies and contact correction}
\label{sec:theory-pathologies}

\setlength{\intextsep}{5pt}
\begin{wrapfigure}{R}{0.5\linewidth}
\centering
    \includegraphics[width=\linewidth]{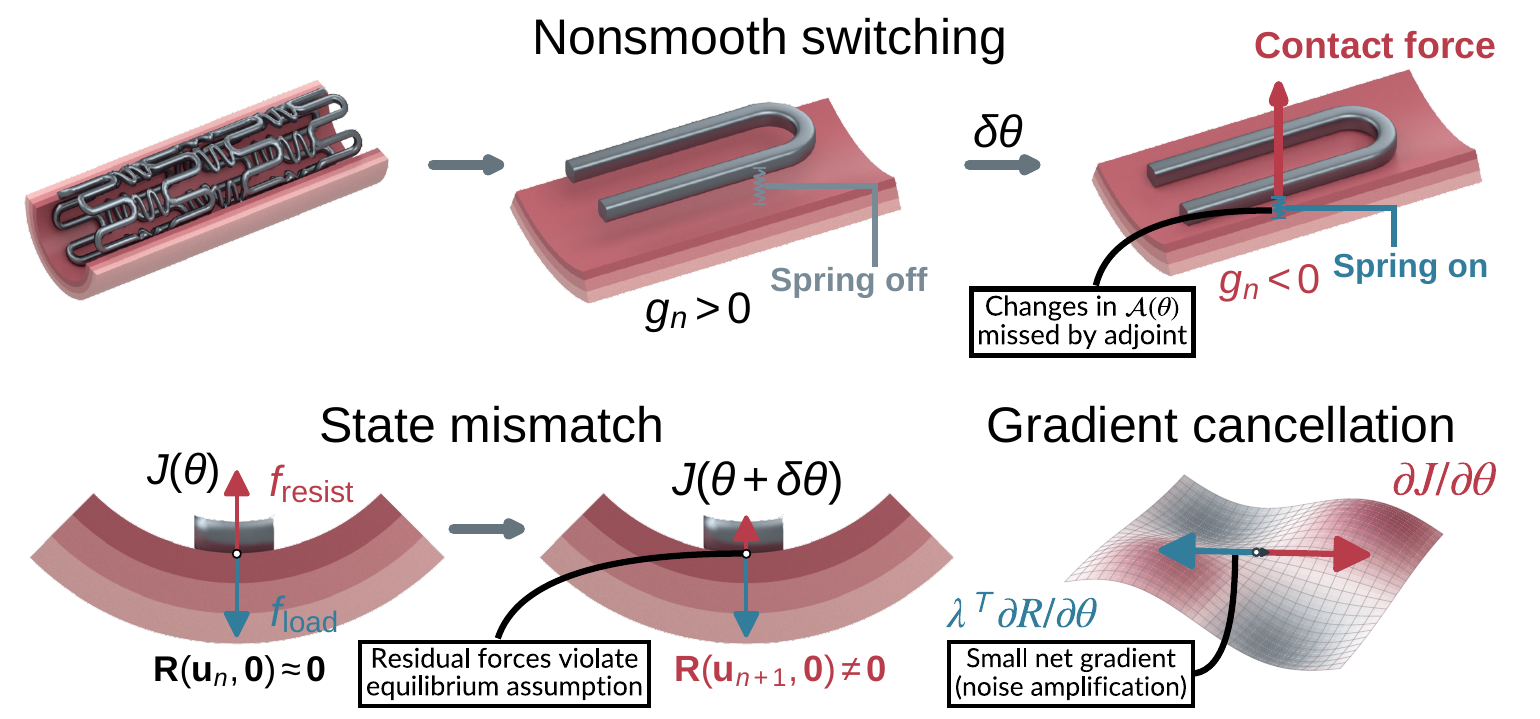}
    \caption{\textbf{Three intrinsic mechanisms limit gradient reliability.} Schematics show contact switching, force imbalance, and cancellation between direct and adjoint contributions.}
    \label{fig:gradient-pathologies}
\end{wrapfigure}

We identify three gradient pathologies that affect the reliability of the computed sensitivities (Figure~\ref{fig:gradient-pathologies}). \textbf{Nonsmooth switching.} The stiff spring contact penalty $\langle{-g_n}\rangle$ is abruptly activated when the gap between a strut and the wall falls below a minimum threshold. The penalty can thus jump discretely across perturbations that alter the active contact set $\mathcal{A}(\boldsymbol{\theta})$ (Appendix~\ref{app:formulation-contact}). Meanwhile, the adjoint trick treats the contact set as fixed, producing systematic errors for perturbations that activate or release contact. Smoothing the activation in the adjoint over a narrow gap interval can reduce the mismatch. \textbf{State mismatch.} Perturbing a material parameter affects the current state indirectly through the entire history of plastic deformation. The adjoint treats the plastic history and force residual $\mathbf{f}=\mathbf{R}(\mathbf{u},\mathbf{0})$ as fixed. Due to velocity-dependent contact, the recorded state is not in static equilibrium, and the adjoint trick thus produces errors in gradients for parameters that alter the force balance, like balloon pressure. \textbf{Gradient cancellation.} The total derivative is the sum of the direct term $\partial J/\partial\boldsymbol{\theta}$ and the adjoint term $-\boldsymbol{\lambda}^{\top}\partial\mathbf{R}/\partial\boldsymbol{\theta}$. If they have opposite signs, small errors in either term are amplified by the remaining difference. Appendix~\ref{app:pathologies} further examines finite-difference reference accuracy, fiber recruitment, and the effects of near-incompressibility on conditioning and discretization.

\section{Experiments}
\label{sec:experiments}

Appendix~\ref{app:forward-validation} validates the simulator against \citet{conway2012computational}: peak lumen area agrees within $11\%$ at the calibrated $857$~kPa; recoil matches without plaque. We use this validated forward model for all subsequent experiments.

\subsection{Testbeds and protocol}
\label{sec:exp-setup}

\textbf{T1} is contact-free: a uniaxial 8 element HGO cube (closed-form solution) and a pressurized HGO tube. \textbf{T2} is the benchmark's three-layer vessel, with or without $60\%$ eccentric stenosis \citep{ohara2008eccentric,conway2012computational}. \textbf{T3} is a calcified coronary segment reconstructed, following \citet{kadry2021platform}, from one patient's anonymized clinical intravascular optical coherence tomography (OCT) data, provided by HeartFlow Inc.\ under IRB approval.

By default, T2 and T3 deploy a balloon-expanded stent (Appendix~\ref{app:formulation-stent}; crimped smaller for T3) with pressure ramping to $857$~kPa at $t=1.5$~s (peak) and to zero at $3.0$~s (Appendix~\ref{app:hyperparams}). Wall constants follow Table~\ref{tab:conway-wall} except $\kappa=0.2$. The T2 reference is the stenosed vessel. Outcomes $\mathcal{J}$ are throat lumen area and volume-mean von Mises stress in each layer and the plaque-touching intima (interface stress). Parameters are each layer's $c$, $k_1$, $k_2$, $\kappa$, $D$ and plaque modulus $G$, with sensitivities $S_\theta=(\theta/\mathcal{J})\,\partial\mathcal{J}/\partial\theta$.

\subsection{Gradient accuracy}
\label{sec:exp-correctness}

\begin{figure*}[h]
    \centering
    \includegraphics[width=\textwidth]{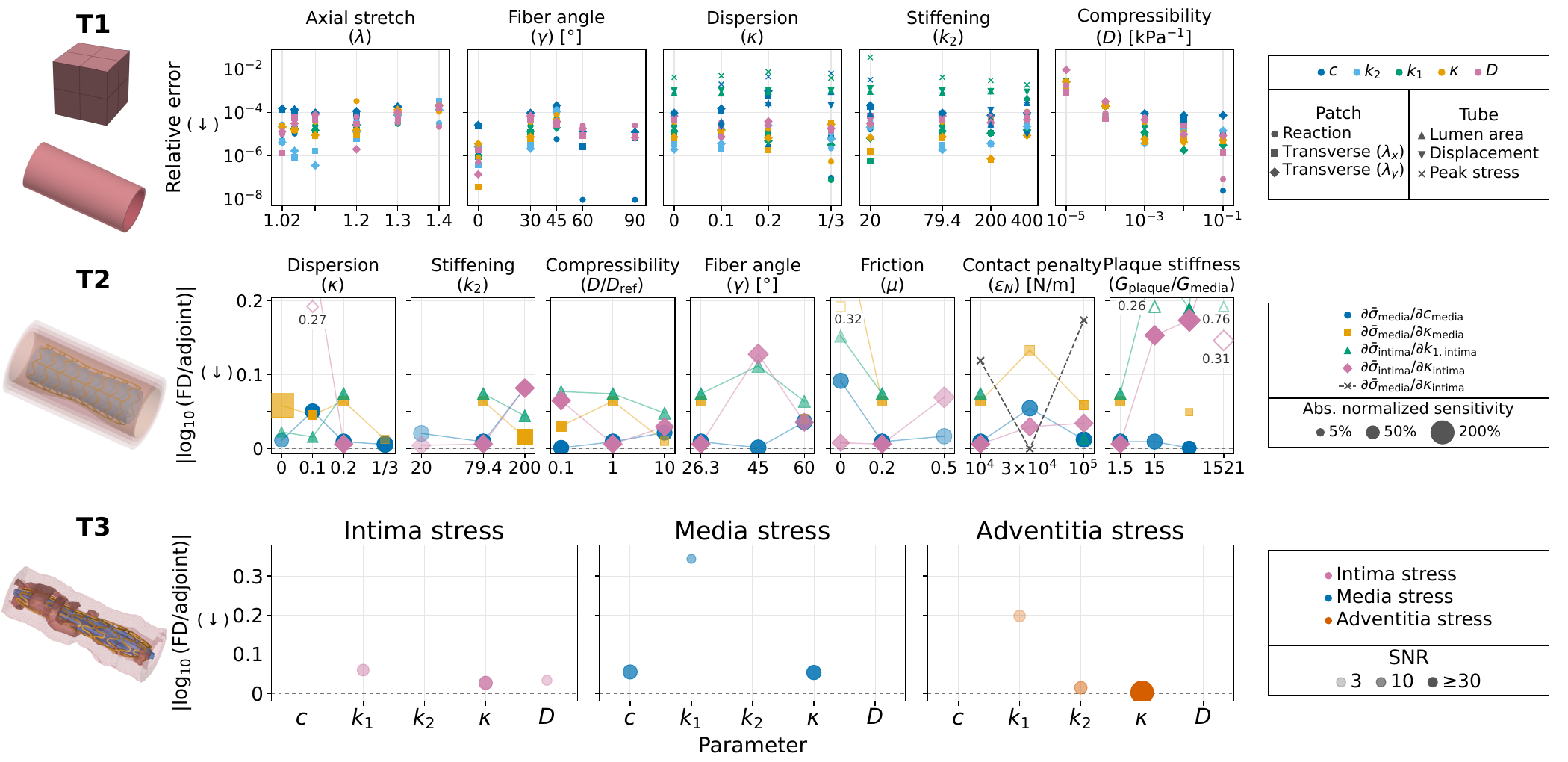}
    \caption{\textbf{Adjoint gradient accuracy before backward-pass treatments.} T1: relative error against closed-form (cube, solid) or finite-difference (tube, dashed) derivatives. T2 (legend pairs), T3 (layer stresses against own parameters): $|\log_{10}(\mathrm{FD}/\mathrm{adjoint})|$ at peak pressure; marker area $|S_\theta|$, opacity SNR. %
    Lower values indicate closer alignment.}
    \label{fig:gradient-accuracy}
\end{figure*}

We compare the unmodified adjoint against central finite difference perturbations ($\pm5\%$, $\pm0.02$ for $\kappa$), holding the contact set and plastic history frozen (Figure ~\ref{fig:gradient-accuracy}). Identical deployments differ slightly due to single-precision atomic sums amplified by discrete fiber recruitment and strut contact. We plot an FD-adjoint gradient pair only when its observed or adjoint-predicted change exceeds three repeat standard deviations (Signal-to-Noise ratio (SNR): observed change in standard deviations). Missing markers fall below this threshold, except five rejected pairs. (Appendix~\ref{app:gradient-validation}).

On T1, cube errors stay below $10^{-3}$ except when near-incompressibility ($D=10^{-5}$~kPa$^{-1}$) makes the force residual fail to be zero, as required by the adjoint. Tube errors stay within $0.13\%$ (lumen area, displacement) and $1\%$ (peak stress), except where direct and adjoint terms nearly cancel.(Tables~\ref{tab:t1-cube},~\ref{tab:t1-tube}).

On T2, all but two plotted own-layer FD/adjoint ratios lie within $0.7$--$1.3$ (media $c$: $0.81$--$1.21$) except at stiffer plaques, where the adjoint under-predicts intima-stress sensitivities ($1.4$--$2.0$ for $\kappa$, $1.5$--$5.8$ for $k_1$), largely unrepaired by contact activation (Section~\ref{sec:exp-ablations}). The recurring defect is that the adjoint over-predicts the mean-media-stress sensitivity to intima $\kappa$ at six stenosed points ($0.67$--$0.84$; Tables~\ref{tab:t2-points},~\ref{tab:t2-contact}). This is addressed in Section ~\ref{sec:exp-ablations}.

On T3, seven of nine resolved ratios lie within $0.87$--$1.13$; the weakly resolved media and adventitia $k_1$ give $2.21$ and $0.63$, and the cross-layer ratio is $0.39$ (SNR $=7$). Appendix~\ref{app:adjoint-error-sources} relates these discrepancies to the observed gradient pathologies of nonsmooth switching, state mismatch, and gradient cancellation.

\subsection{Comparison of adjoint formulation against traditional methods}
\label{sec:exp-adjoint}

\setlength{\intextsep}{5pt}
\begin{wrapfigure}[37]{R}{0.48\linewidth}
\centering
    \includegraphics[width=\linewidth]{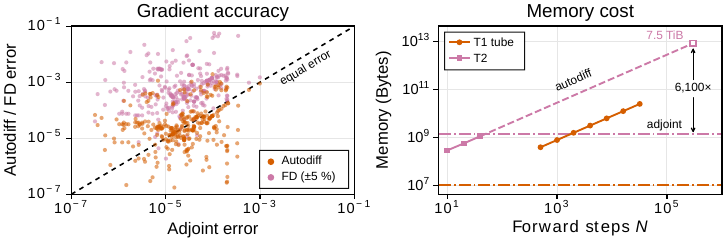}
    \caption{\textbf{The adjoint is as accurate as autodiff, and its memory does not grow with simulation size.} Left: relative error of each derivative against the closed-form ($327$ derivative values, summarized in Table~\ref{tab:t1-methods}). Points above the dashed line are less accurate than the adjoint. Right: GPU memory required for differentiation. Warp tape memory is measured on the T1 tube and T2, and extrapolated linearly to the full deployment. Adjoint memory does not depend on $N$.}
\label{fig:adj-vs-unrolled}
\vspace{6pt}
    \includegraphics[width=\linewidth]{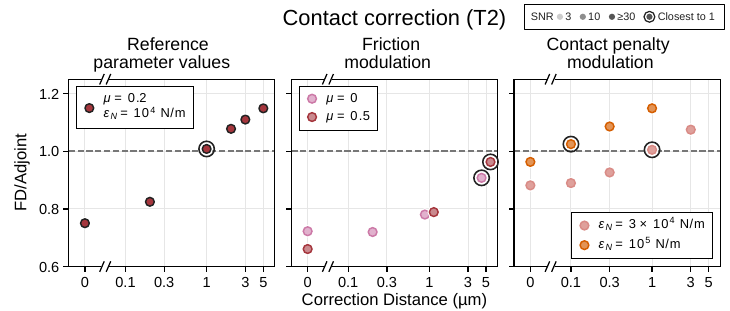}
    \caption{\textbf{Contact activation improves adjoint accuracy, but no single correction distance suits every contact setting.} FD/adjoint agreement for sensitivity of mean media stress to intima $\kappa$ on stenosed T2 at peak inflation pressure, against the correction distance $w$ (the width of the contact-activation sigmoid; $w=0$: no correction). The panels vary friction $\mu$ and contact penalty $\epsilon_N$ around the reference values. The dashed line marks exact agreement, and circles mark the distance closest to it in each series.}
\label{fig:contact-activation}
\end{wrapfigure}

We compare the adjoint against Warp's unrolled differentiation (tape) through every integration step on the T1 tube (Figure~\ref{fig:adj-vs-unrolled}). The memory required by autodiff grows $0.73$~MiB (tube) and $26.3$~MiB (T2) per step, an extrapolated $7.5$~TiB per $300{,}000$-step deployment. The adjoint is significantly more efficient. It needs only $10$~MiB (coarser tube) and $1.26$~GiB (one T2 load step).

\subsection{Ablations}
\label{sec:exp-ablations}

The \textbf{contact activation} correction introduces a width \(w\) over which the adjoint's contact stiffness transitions as a sigmoid between inactive and active contact. (Appendix~\ref{app:treatments}). Figure~\ref{fig:contact-activation} shows that this correction improves agreement with finite differences for the sensitivity of mean media stress to intima dispersion \(\kappa\), with little change to the already accurate media $c$ sensitivity. The improvement depends on smoothing width, contact stiffness, friction, and deployment stage: no single width improves every condition. Our results indicate that contact activation is a contributor to the gradient discrepancy, but it does not fully explain the mismatch.

The second treatment is an adjoint that accounts for \textbf{plastic strain accumulation}. Plastic strain accumulates along the loading path, so changing the yield stress changes the strain carried into every later load step. The endpoint adjoint treats the recorded plastic strain as fixed and misses this effect; we introduce a chained adjoint that differentiates through each step's plastic update in order to include it. To isolate the effect, we stretch an elastoplastic block by $30\%$, unload it to $18\%$, and differentiate its lateral displacement with respect to the yield stress. The endpoint adjoint predicts almost no sensitivity ($-0.01\times$ the finite difference), whereas the chained adjoint recovers its sign and magnitude ($1.25\times$). On T2, chaining changes the media sensitivities by less than $1\%$. After deflation it reduces the plaque-modulus FD/adjoint ratio from $4.5$ to $2.6$ (SNR $=6$) but moves the cross-layer ratio (mean media stress to intima $\kappa$) away from agreement, from $0.92$ to $0.84$. Neither adjoint recovers the stent's yield-stress sensitivity: at $t=1.0$~s both capture only $10$--$24\%$ of the finite difference (SNR $\geq120$). We attribute this to the recorded stent's force imbalance (Section~\ref{sec:exp-clinical}) and to the adjoint holding fixed which stent elements have yielded, the plastic analogue of the frozen contact set. The remaining experiments use the endpoint adjoint without either treatment.

\subsection{Material sensitivity}
\label{sec:exp-sensitivity}

\begin{figure}[t]
    \centering
    \includegraphics[width=0.95\linewidth]{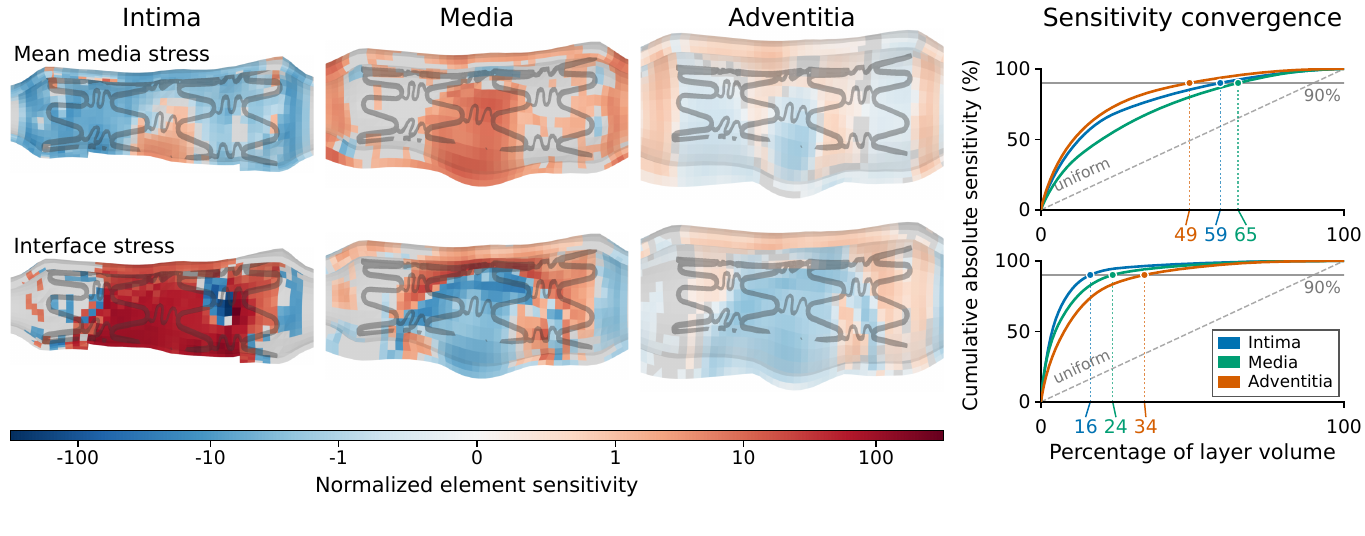}
    \caption{\textbf{Most of the sensitivity to stiffness is concentrated in a few tissue elements.} T3 artery shown at peak pressure. Each tissue element $e$ is given its own ground-substance stiffness $c_e$; its sensitivity $s_e = c_e\,\partial J/\partial c_e$ is the change in the outcome $J$ when only that element is stiffened, per relative change in $c_e$. All $s_e$ for one outcome come from a single adjoint solve. Red/blue: stiffening the element raises/lowers $J$); Gray: elements outside the smallest set carrying $90\%$ of the layer's $\sum|s_e|$. Right: cumulative share of the layer's sensitivity against layer volume, elements ranked by sensitivity per unit volume.}
    \label{fig:element-sensitivity}
    \vspace{-6pt}
\end{figure}

\setlength{\intextsep}{5pt}
\begin{wrapfigure}[31]{R}{0.4\linewidth}
\centering
\includegraphics[width=\linewidth]{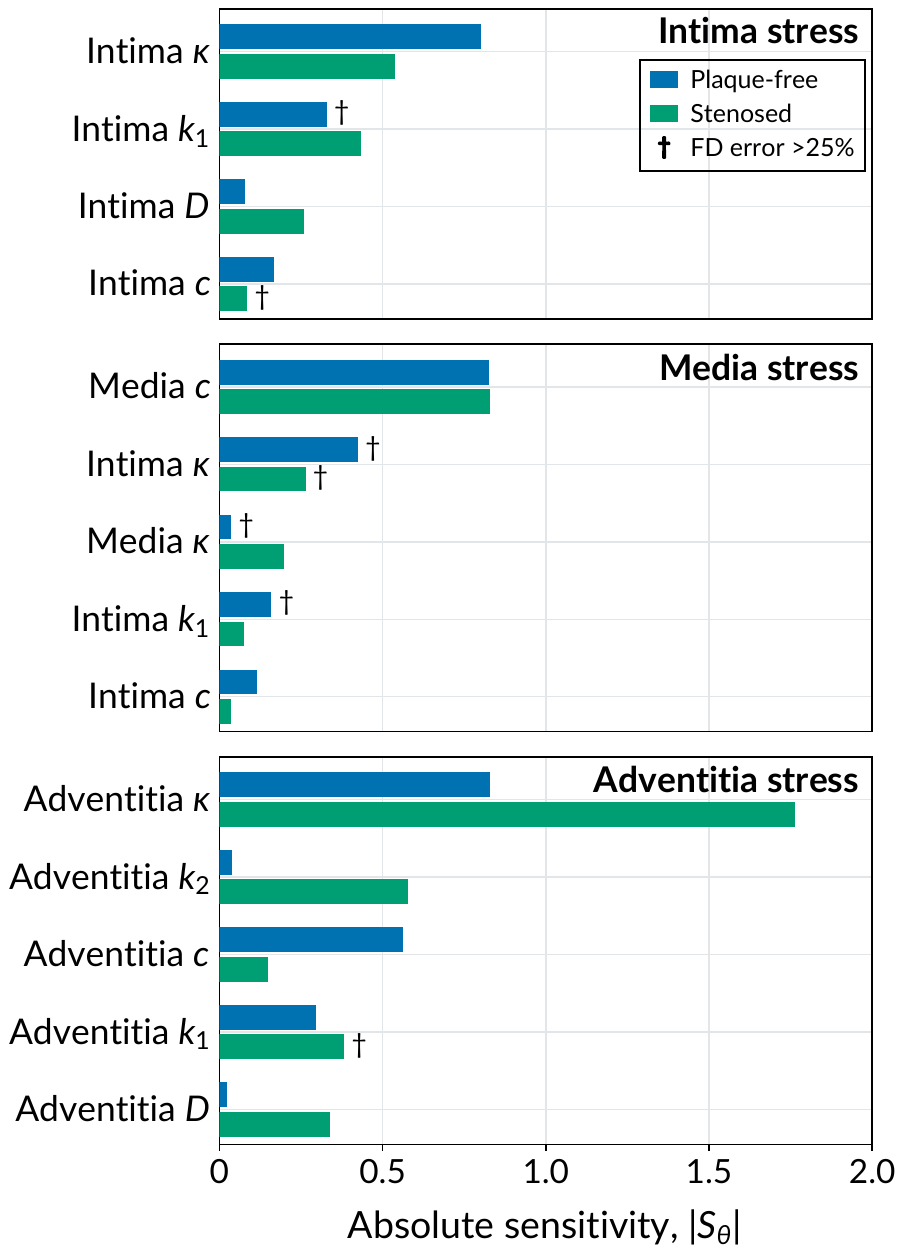}
\caption{\textbf{Material sensitivities differ between plaque-free and stenosed vessels.} Absolute endpoint-adjoint sensitivities $|S_\theta|$ of each layer's mean stress at peak pressure in plaque-free and stenosed T2. The daggers indicate adjoint error above $25\%$ of FD.}
\label{fig:material-sensitivity}
\end{wrapfigure}

Calcification changes how a stent loads the arterial wall, and may therefore affect which tissue properties control wall stress. This sensitivity is difficult to measure in a real patient's arteries, and computationally expensive with finite differencing. The adjoint trick makes it much easier, requiring only one backward solve per outcome yields the sensitivity to every material parameter. We compare absolute sensitivities $S_\theta$ between the plaque-free and stenosed T2 vessels at peak pressure.

On both stenosed and unstenosed T2 vessels (Figure~\ref{fig:material-sensitivity}), the same parameters dominate sensitivity: stiffness $c$ for the media, and fiber dispersion $\kappa$ for the intima and adventitia. Stenosis doubles the sensitivity of adventitia stress to adventitia $\kappa$ ($|S_\theta|$: $0.83\to1.76$) and increases that of media stress to media $\kappa$ ($0.04\to0.20$). Finite differences confirm both stenosed values.

Next, we ask how stiffness is localized within each layer of a wall. To identify which parts of each layer control wall stress, we treat the ground-substance stiffness $c_e$ of each of the $26{,}320$ T3 tissue elements as a separate parameter. Its sensitivity $s_e=c_e\,\partial\mathcal{J}/\partial c_e$ is the change in $\mathcal{J}$ per relative change in $c_e$. Finite differences would require two deployments per element; the adjoint provides all $s_e$ for one outcome from a single backward solve. (Figure~\ref{fig:element-sensitivity}).

Media elements account for $90\%$ of the summed sensitivity of the mean media stress, and each has $s_e>0$. Intima elements account for $9\%$, $81\%$ of it from elements with $s_e<0$, so stiffening most of the intima lowers the media stress. For the interface stress, intima elements account for $67\%$ and media elements for $27\%$. The sensitivity is concentrated: in each layer, the smallest set of elements carrying $90\%$ of the summed $|s_e|$ occupies only $25$--$49\%$ of the layer's volume, against $90\%$ if sensitivity were spread evenly.

\subsection{Identification and Optimization}
\label{sec:exp-clinical}

\begin{wrapfigure}[25]{R}{0.65\linewidth}
\centering
    \includegraphics[width=1\linewidth]{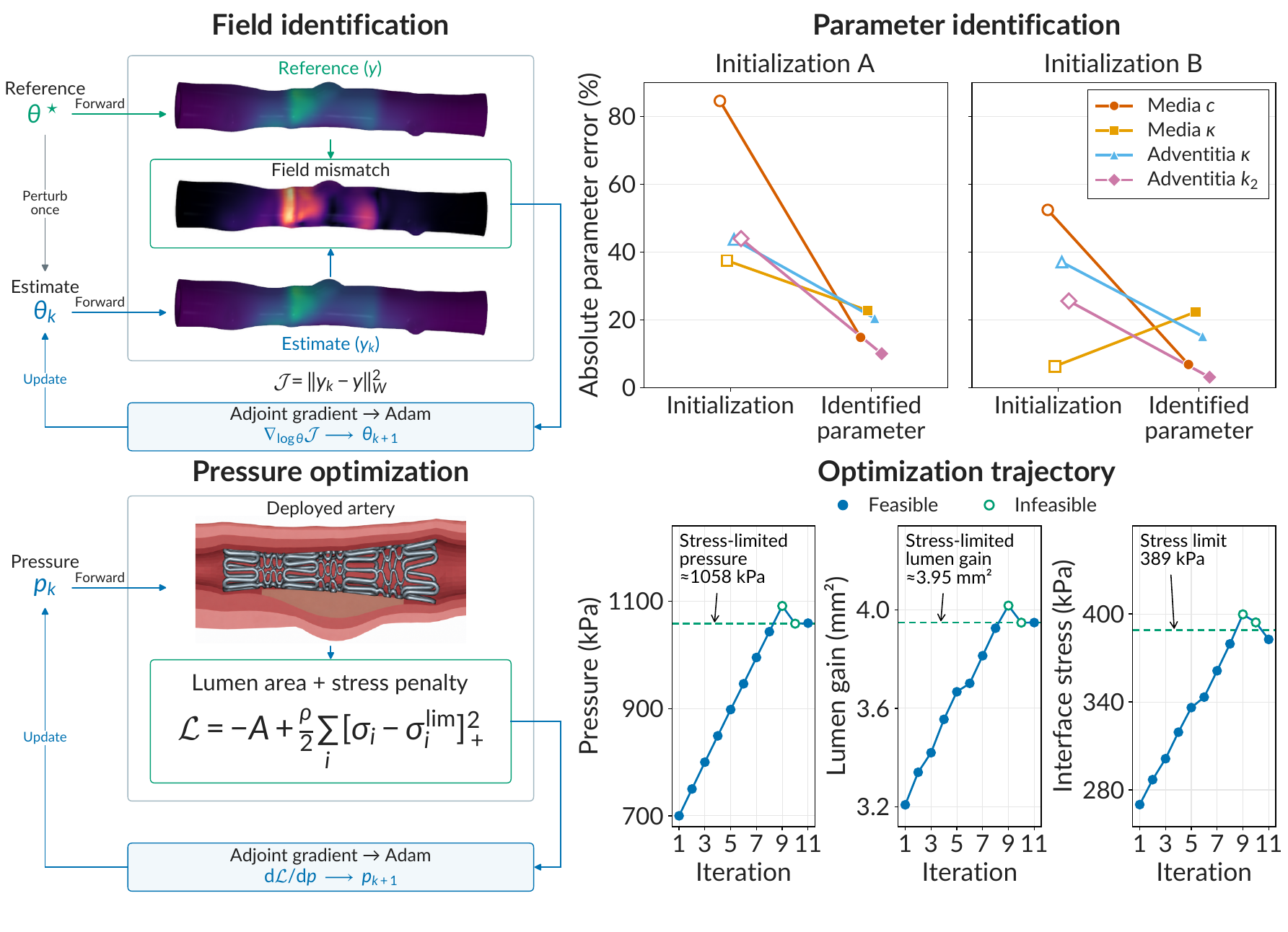}
    \caption{\textbf{Material identification (top) and stress-constrained pressure optimization (bottom).} Top right: parameter error of two fits (diamonds: start; bars: geometric mean of the last three of eight evaluations). Bottom: evaluations 1--9 are Adam, 10 the adjoint Newton step, 11 a second deployment $1$~kPa higher. Lumen gain is over the pre-contact throat lumen ($2.63$~mm$^2$).}
    \label{fig:identification_optimization}
\end{wrapfigure}

We display the capabilities of the adjoint by solving two inverse problems: parameter identification and constrained optimization, visualized in Figure ~\ref{fig:identification_optimization}. Identification takes as input the displacement of all $30{,}976$ T3 tissue nodes at peak pressure, averaged over four deployments at the reference parameters. We fit four parameters (media $c$ and $\kappa$, adventitia $\kappa$ and $k_2$) by minimizing the mean squared nodal misfit $\mathcal{J}$ with Adam on $\log\boldsymbol{\theta}$ and endpoint-adjoint gradients, for eight evaluations from each of two random starts $6$--$85\%$ off (Figure~\ref{fig:identification_optimization}).

\setlength{\intextsep}{5pt}

A new deployment at the reference parameters typically differs from the averaged observation by $\mathcal{J} \approx 3.5\times10^{-6}$ mm$^2$, and both fits converge to $0.8$--$1.3$ times this repeat baseline. But the estimates, geometric means of the last three evaluations, remain $3$--$23\%$ off, and media $\kappa$ is predicted $22\%$ low from both starts. The difference affects forward prediction of media stress, inflating it $\times19\%$. Matching the load state's displacement field therefore does not sufficiently constrain the parameters that set wall stress, even without imaging noise. 

Stent deployment requires an inflation pressure that trades lumen gain against wall injury. The optimization problem on T3 is to maximize the lumen area $\mathcal{L}(p)$ at the stenosis throat while keeping the mean media stress and the plaque-adjacent intima stress, $C_i(p)$, below bounds $\bar{C}_i$ that is set at $1.2\times$ their values at the calibrated reference pressure.

We solved the problem with a quadratic penalty and Adam on the scaled pressure (Fig.~\ref{fig:identification_optimization}). From a low starting pressure the method climbs in the direction prescribed by the adjoint until it crosses the interface-stress bound. The lumen and stress both increase monotonically with pressure, so the optimum is the root of the binding constraint, $g(p) = C_1(p) - \bar{C}_1 = 0$. We therefore follow the ascent up to its first infeasible evaluation, which brackets the root, and take a Newton step on $g$ with the adjoint slope $\mathrm{d}C_1/\mathrm{d}p$, followed by one replicate as a noise check (Fig.~\ref{fig:identification_optimization}). The constraint boundary is found in a single Newton step. The precision is limited by the scatter of the interface stress between identical deployments by tens of kilopascals. The stress-limited pressure, $1058\pm20$~kPa ($\sigma_1$ noise: $2\%$ about its trend vs.\ $0.4\%$ between repeats), is $23\%$ above $857$~kPa, raising $A$ from $6.22$ to $6.58$~mm$^2$ for two further deployments and one adjoint solve.

\section{Conclusions}
We present DiffStent, a validated GPU-accelerated differentiable simulator for anisotropic arterial contact during stent deployment. For efficient gradient computation, we contribute an adjoint gradient method based on the implicit function theorem, which outperforms reverse-mode autodiff both in accuracy and memory cost. Our method overcomes the pathologies of the adjoint that stem from cross-linking anisotropic fiber constitutive models, namely the treatment of switch-like parameters as fixed and cancellation of direct and adjoint gradient terms. We show improvement in sensitivities across constitutive parameters through the backward-pass contact surrogate while preserving the forward mechanics. The clinical relevance is demonstrated by computing the effects of calcification on sensitivity and the localization of stress sensitivity in each arterial layer. Finally we use DiffStent to solve an inverse problems of parameter identification via constrained optimization. Our work thus provides a practical foundation for gradient-based inference and optimization in computational cardiovascular mechanics, with the potential to enable patient-specific procedural planning in interventional settings.

\subsection*{AI use statement}

In this work, we used Claude Code to assist with implementing methods, executing experiments, and analyzing and interpreting results, including implementing the synthetic data generation procedures for T1 and T2. We have not used generative AI tools for research ideation, refinement of hypotheses, design or critique of methodology or experiments, or other tasks requiring disclosure. Additionally, we used generative AI tools to draft and edit parts of the paper text, including condensing captions and helping to arrange figures in \LaTeX. We used Claude Code to draft figures. We reviewed all AI-assisted output: we checked and tested all code before using it for any experiment, and verified text and figures against our methods and results. We take responsibility for the final content of this work, including text, claims, or artifacts produced with the aid of generative AI.

\subsection*{Ethics statement}

This work uses a coronary model derived from optical coherence tomography (OCT) images of a single patient (T3). 
HeartFlow Inc. provided the anonymized imaging data, whose use was authorized under IRB protocol 2022P001679, “Personalized lesion modification optimizes atherosclerosis intervention,” with exempt status.

\subsubsection*{Acknowledgments}
We thank HeartFlow Inc. for providing the patient data used in this study. This work was supported by the American Heart Association under award 25CDA1452999. Computing resources were provided by the NSF-ACCESS award MDE250010. This support is gratefully acknowledged.

\bibliography{iclr2027_conference}
\bibliographystyle{iclr2027_conference}

\newpage
\appendix

\section{Appendix}
\label{app:overview}

The appendix provides additional formulations, validation results, and analyses of gradient reliability to support the main manuscript, organized as follows:

\begin{itemize}[leftmargin=1.6em, itemsep=2pt, topsep=3pt, parsep=0pt]
  \item \textbf{Section~\ref{app:forward-validation} (Forward-model validation):} Compares simulated stent expansion and recoil with the published benchmarks and describes pressure calibration and discretization effects.
  \item \textbf{Section~\ref{app:formulation} (Continuous and discrete formulation):} Specifies fiber dispersion, the weak form and load stepping, contact and friction, and the stent and balloon models.
  \item \textbf{Section~\ref{app:derivations} (Derivation of the stress and consistent tangent):} Presents the stress and fiber tangents derived from the constitutive energy.
  \item \textbf{Section~\ref{app:adjoint} (Adjoint derivation and computational cost):} Describes history dependence, tangent assembly, and the computational cost of adjoint sensitivities relative to finite differences and unrolled differentiation.
  \item \textbf{Section~\ref{app:pathologies} (Supplementary gradient analysis):} Examines finite-difference and complex-step references, fiber recruitment, conditioning, and volumetric locking.
  \item \textbf{Section~\ref{app:treatments} (Backward-pass treatment specifications):} Specifies the chained plastic-history adjoint and the contact activation correction.
  \item \textbf{Section~\ref{app:hyperparams} (Solver tolerances, mesh convergence, and hyperparameters):} Documents the testbeds, repeat protocol, forward and adjoint solvers, and mesh-refinement studies.
  \item \textbf{Section~\ref{app:gradient-validation} (Additional gradient-validation results):} Reports detailed gradient comparisons on T1--T3, finite-difference resolvability, and baseline accuracy and cost.
  \item \textbf{Section~\ref{app:adjoint-error-sources} (Sources of adjoint error):} Relates the observed discrepancies to nonsmooth switching, state mismatch, and gradient cancellation.
  \item \textbf{Section~\ref{app:unrolled-validation} (Parameter-wise gradient errors):} Compares peak-stress sensitivities from the equilibrium adjoint and unrolled autodiff against a relaxed adjoint reference.
  \item \textbf{Section~\ref{app:limitationsandfuture} (Limitations and Future Work):} Outlines extensions to retrospective clinical cohorts and amortized inference of mechanical property maps from imaging.
\end{itemize}

\section{Forward-model validation}
\label{app:forward-validation}

The simulator that every experiment differentiates is first run, without any modification, on the coronary stent implantation test-bed of \citet{conway2012computational}, which reports lumen area at maximum inflation and after deflation, and the resulting elastic recoil, for a closed-cell stent deployed in a straight vessel at three stenosis levels (Figure~9 in \citet{conway2012computational}).

\paragraph{Benchmark.} A 3 mm coronary with a 0.5 mm wall in three equal layers (intima, media, adventitia), each an HGO solid with the parameters given in Table~\ref{tab:conway-wall}, with a 10 mm eccentric semi-ellipsoidal plaque reducing the lumen area by $50\%$ (SAL50) or $60\%$ (SAL60), or no plaque. The plaque is third-order Mooney--Rivlin solid with a $0.4$ MPa yield stress. Stent A is the CYPHER-like design of Appendix~\ref{app:formulation-stent}, expanded by a folded $850$ MPa nylon balloon with $\mu = 0.2$ on every contacting pair, over a 3 s linear pressure ramp followed by a 3 s deflation.

\begin{table}[h]
\centering
\small
\caption{Arterial-wall constants of the benchmark (Table~1 of \citet{conway2012computational}), used verbatim: $\kappa = 0$ in every layer, $K = 2/D$, and each layer one third of the 0.5 mm wall.}
\label{tab:conway-wall}
\setlength{\tabcolsep}{6pt}
\begin{tabular}{@{}lrrrrrr@{}}
\toprule
layer & thickness (mm) & $c$ (kPa) & $k_1$ (kPa) & $k_2$ & $D$ (kPa$^{-1}$) & $K$ (kPa) \\
\midrule
intima & 0.167 & 46.8 & 9993 & 80.2 & $4.30\times10^{-4}$ & 4647 \\
media & 0.167 & 19.0 & 21.0 & 79.4 & $1.06\times10^{-3}$ & 1889 \\
adventitia & 0.167 & 6.6 & 118.5 & 373.4 & $3.04\times10^{-3}$ & 658 \\
\bottomrule
\end{tabular}
\end{table}

\paragraph{Discretization.} The tissue is a structured mesh of $23{,}040$ trilinear hexahedra ($29{,}040$ nodes) integrated with $2\times2\times2$ Gauss points, built so that the minimum cell thickness is $0.074$ mm; the stent is about $58$k linear tetrahedra. The same deployments were also run with the tissue as $138$k linear tetrahedra at the same node budget. Before the deployment, each element was gated on its own: the hexahedral J2 return map against the tetrahedral one on a compressed cube (reaction force within $1\%$, equivalent plastic strain ratio $0.98$); the hexahedral HGO cube against the closed-form uniaxial solution (linear $2.7\%$, neo-Hookean $0.0\%$, HGO at most $13.8\%$ at $20\%$ strain); and the constrained-end arterial strip of \citep{gasser2005hyperelastic}, whose ratio of crossings between the two dispersion cases is reproduced to $-2.6\%$ ($1.654$ against $1.697$).

\paragraph{Pressure calibration.} The reference fixes its peak pressure at $1.1$ MPa as the pressure at which their stent, expanded without a vessel, reaches a 3.0 mm inner diameter. Measured the same way (stent and balloon only, wall contact disabled), our stent crosses 3.00 mm at $0.857$ MPa and is balloon-limited above about $0.85$ MPa (Table~\ref{tab:free-stent}). The headline comparison therefore uses $0.857$ MPa; the rows at the reference's own $1.1$ MPa are reported alongside.

\begin{table}[h]
\centering
\small
\caption{Free expansion of stent A (no vessel) at t = 3 s}
\label{tab:free-stent}
\begin{tabular}{@{}lcccccc@{}}
\toprule
pressure (MPa) & 0.66 & 0.75 & 0.81 & 0.857 & 0.92 & 1.10 \\
\midrule
inner diameter (mm) & 2.51 & 2.86 & 2.97 & 3.00 & 3.04 & 3.07 \\
\bottomrule
\end{tabular}
\end{table}

\paragraph{Readout.} Lumen area is the tissue-bounded cross-section minus the strut metal cut by the section, normalized by the open 3 mm vessel with the stent expanded to a 3.00 mm inner diameter ($\pi\cdot1.5^2 - 0.242$ mm$^2$), measured at the plaque throat (within $1.5$ mm of the mid-section) and taken as the maximum over the whole cycle, since the yielding stent and plaque keep expanding $0.3$--$0.8$ s into deflation. The reference normalizes by the bare $\pi\cdot1.5^2$ at the mid-section; the two ratios differ by under $1\%$. Recoil is $(A_{\max} - A_{\mathrm{def}})/A_{\max}$.

\paragraph{Result.} Table~\ref{tab:conway} lists the maximum area, the deflated area, and the recoil for Stent A in the straight vessel of \citet{conway2012computational}. The catheter, guidewire, and the reference's mass-proportional damping are not modelled. Three observations follow. First, in the plaque-free vessel the deployed area and the recoil ($5.1$--$5.9\%$ against $5.7\%$) match the reference, and hexahedra and tetrahedra agree to three digits: the wall itself does not lock. Second, every difference between the two elements in the stenosed vessels is in the plaque, where linear tetrahedra lock at $K/c \approx 450$ and under-expand the lumen by up to $11\%$. This locking is discussed in Appendix~\ref{app:pathologies-locking} and is the reason the tissue is later discretized with hexahedra. Third, at the calibrated pressure the plaque-free and SAL50 maxima sit within $+7\%$ and $-4\%$ of the reference while SAL60 under-expands by $11\%$ (it matches at $1.1$ MPa): our stenosed lumen is more sensitive to pressure than the reference's. The open discrepancy is the pattern of recoil, $11\%$ at both stenosis levels here against $8.5\%$ at SAL50 and $18\%$ at SAL60 in Figure~9 of \citet{conway2012computational}. We suspect that this discrepancy might be due to one-point reduced time integration used in the reference.

\begin{table}[h]
\centering
\small
\caption{Maximum lumen area / deflated lumen area / recoil in an idealized straight vessel}
\label{tab:conway}
\setlength{\tabcolsep}{4pt}
\begin{tabular}{@{}lllll@{}}
\toprule
vessel & hexahedra, 0.857 MPa & hexahedra, 1.1 MPa & tetrahedra, 1.1 MPa & \citet{conway2012computational} \\
\midrule
plaque-free & 1.184 / 1.124 / 5.1\% & 1.246 / 1.172 / 5.9\% & 1.244 / 1.152 / 7.4\% & 1.102 / 1.039 / 5.7\% \\
SAL50 & 1.042 / 0.923 / 11.4\% & 1.176 / 1.041 / 11.5\% & 1.144 / 0.985 / 13.9\% & 1.087 / 1.022 / 8.5\% \\
SAL60 & 0.879 / 0.783 / 10.9\% & 1.100 / 0.956 / 13.0\% & 0.994 / 0.855 / 14.0\% & 0.985 / 0.805 / 18.2\% \\
\bottomrule
\end{tabular}
\end{table}

\section{Continuous and discrete formulation}
\label{app:formulation}

\subsection{Fiber dispersion}
\label{app:formulation-dispersion}

The mean fiber directions of Section~\ref{sec:theory-constitutive} are $\mathbf{a}_{0i} = \cos\gamma\,\mathbf{e}_{\theta} \pm \sin\gamma\,\mathbf{e}_{z}$, $i = 1,2$: the two collagen families wind helically in the tangential plane at $\pm\gamma$ to the circumferential direction, with $\gamma$ varying transmurally across the intima, media, and adventitia. Fibers are not perfectly aligned about $\mathbf{a}_{0i}$. Following \citep{gasser2005hyperelastic}, we model the orientation distribution as a transversely isotropic von Mises density $\rho(\Theta)$ about the mean direction and condense it into the scalar dispersion
\begin{equation}
\kappa = \tfrac{1}{4}\int_{0}^{\pi}\rho(\Theta)\sin^{3}\Theta\,\mathrm{d}\Theta, \quad \kappa \in [0,\tfrac{1}{3}],
\label{eq:kappa-integral}
\end{equation}
which enters the energy only through the generalized structure tensor $\mathbf{H}_i = \kappa\mathbf{I} + (1-3\kappa)\,\mathbf{a}_{0i}\otimes\mathbf{a}_{0i}$. On the benchmark of Appendix~\ref{app:forward-validation} every layer carries the published constants with $\kappa = 0$ (perfectly aligned families); $\kappa$ is swept on T1 and T2 only. On the idealized testbeds, the plaque-to-media stiffness ratio is varied; patient-specific calcific inclusions are isotropic elements with $k_1=0$ and elevated $c$.

\subsection{Weak form and load stepping}
\label{app:formulation-weakform}

\begin{figure}[htbp]
\centering
\includegraphics[width=0.5\linewidth]{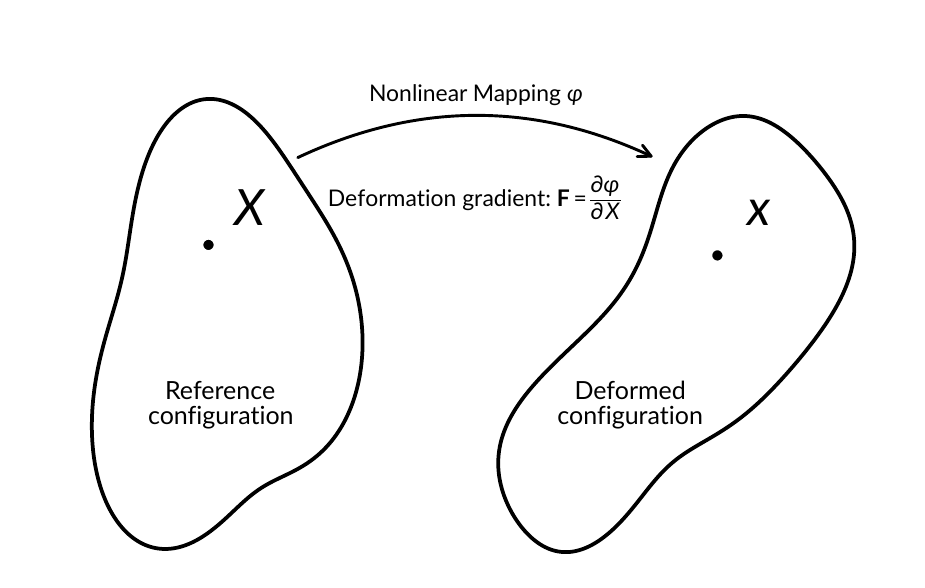}
\caption{\textbf{Nonlinear deformation between reference and current configurations.} The map $\mathbf{x}=\varphi(\mathbf{X})$ relates a material point in reference configuration $\mathbf{X}$ to deformed configuration $\mathbf{x}$, and $\mathbf{F}=\partial\varphi/\partial\mathbf{X}$ defines gradient of deformation at each point in the domain.}
\label{fig:nonlinear-mapping}
\end{figure}

We adopt classical continuum mechanics approach in which the deformation map $\varphi(\mathbf{X})=\mathbf{X}+\mathbf{u}(\mathbf{X})$ takes a material point from the reference configuration to its deformed position. Its local derivative is the deformation gradient $\mathbf{F}=\partial\varphi/\partial\mathbf{X}$ (Figure~\ref{fig:nonlinear-mapping}).

Arterial tissue is nearly incompressible, so we adopt the Flory decomposition $\bar{\mathbf{C}} = J^{-2/3}\mathbf{C}$ of $\mathbf{C} = \mathbf{F}^{\top}\mathbf{F}$ into a volume-preserving part and write the strain energy \eqref{eq:energy} as an additive function of $J$ and $\bar{\mathbf{C}}$. Deployment is quasi-static at physiological rates, so equilibrium reduces to $\nabla_{\mathbf{X}}\!\cdot(\mathbf{F}\mathbf{S}) = \mathbf{0}$ in $\Omega_0$. In the weak form,
\begin{equation}
\int_{\Omega_0}\mathbf{F}\mathbf{S}:\nabla_{\mathbf{X}}\delta\mathbf{u}\,\mathrm{d}V + \delta W_c - \int_{\Gamma_N}\bar{\mathbf{t}}(\beta)\cdot\delta\mathbf{u}\,\mathrm{d}A = 0
\qquad \text{for all admissible } \delta\mathbf{u},
\label{eq:weakform}
\end{equation}
with displacement prescribed on $\Gamma_D$, traction on $\Gamma_N$, and unilateral contact on $\Gamma_C$. The load parameter $\beta$ is the inflation pressure applied to the folded balloon membrane (Appendix~\ref{app:formulation-stent}), ramped over N loading steps during inflation and deflation.

\subsection{Contact and friction}
\label{app:formulation-contact}

Let $g_N(\mathbf{u})$ be the signed gap between two contacting surfaces (balloon and stent, or strut and lumen; both pairs are treated as cross-mesh penalty interfaces). Non-penetration is the Karush--Kuhn--Tucker system $g_N \geq 0$, $t_N \leq 0$, $t_N g_N = 0$, which we regularize by a penalty, softened within a thin band of thickness $h_c$ about the surface, and close with regularized Coulomb friction acting against the tangential relative velocity $\mathbf{v}_T$:
\begin{equation}
t_N = -\epsilon_N\left\langle -g_N\right\rangle,
\qquad
\mathbf{t}_T = -\mu\left|t_N\right|\frac{\mathbf{v}_T}{\max\left(\|\mathbf{v}_T\|, v_\varepsilon\right)},
\qquad
\delta W_c = \int_{\Gamma_C}\left(t_N\,\delta g_N + \mathbf{t}_T\cdot\delta\mathbf{g}_T\right)\mathrm{d}A .
\label{eq:contact}
\end{equation}

Let $\mathcal{P}_C$ be the fixed set of collocation points at which the interface integral in ~\eqref{eq:contact} is evaluated, and $\mathbf{u}^\star(\boldsymbol{\theta})$ the equilibrium state reached at parameters $\boldsymbol{\theta}$. The active contact set is
\begin{equation}
\mathcal{A}(\boldsymbol{\theta})
= \left\{\, p\in\mathcal{P}_C \;:\; g_N^p\big(\mathbf{u}^\star(\boldsymbol{\theta})\big) < 0 \,\right\},
\label{eq:active-set}
\end{equation}
which depends on $\boldsymbol{\theta}$ only through $\mathbf{u}^\star$ and is therefore piecewise constant, changing only when a gap crosses zero. The Macaulay bracket keeps $t_N$ continuous across such a change, but $\partial t_N/\partial g_N$ jumps from $0$ to $\epsilon_N$, so the tangent $\mathbf{K}$ that the adjoint assembles at $\mathbf{u}^\star$ jumps with it.

The friction coefficient is $\mu = 0.2$ on every contacting pair, following the benchmark of Appendix~\ref{app:forward-validation}. Because $\mathbf{t}_T$ is velocity-driven, it vanishes from the equilibrium residual that the adjoint differentiates: friction shapes the configuration the forward pass settles into (which contact pairs are active, and where the struts are held on the lumen), and the adjoint sees that configuration as a frozen active set $\mathcal{A}(\boldsymbol{\theta})$ (\eqref{eq:active-set}). This dependence of the reached configuration on friction contributes to state mismatch when the adjoint differentiates a residual that omits these forces (Section~\ref{sec:theory-pathologies}).

\subsection{Stent and balloon models}
\label{app:formulation-stent}

\begin{wrapfigure}{R}{0.3\linewidth}
    \centering
    \includegraphics[width=\linewidth]{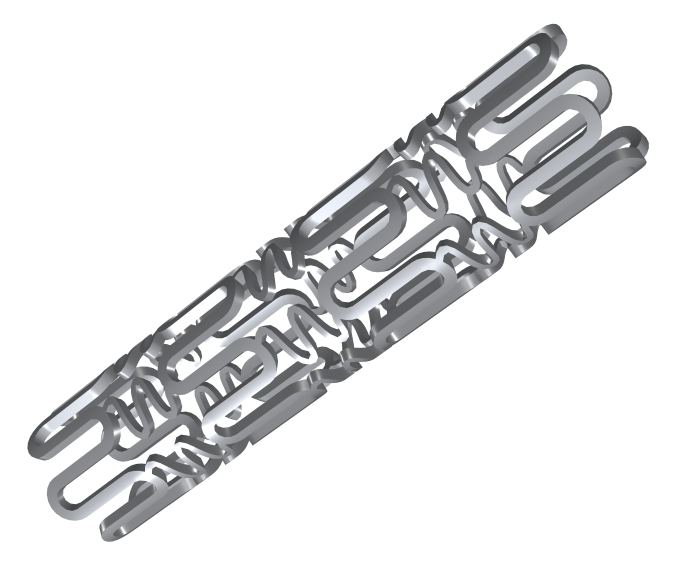}
    \caption{The CYPHER-like closed-cell stent used in T2 and T3: four sinusoidal rings with six peaks each and $0.13 \times 0.14$~mm struts.}
    \label{fig:placeholder}
\end{wrapfigure}
The stent is a closed-cell, sinusoidal-ring design (stent A of the benchmark of Appendix~\ref{app:forward-validation}, generated from the three-parameter CYPHER-like description of \citet{pant2012geometry}), with six peaks per ring, four rings, and a $0.13 \times 0.14$ mm strut section, generated directly in its crimped state at a $1.5$ mm inner diameter, so no crimping stage is simulated for T2. For T3 it is crimped in simulation, by jaws closing from $1.0$ to $0.56$ mm radius, onto a balloon folded more tightly (crimper radius $0.40$ mm), so that it enters the narrower lesion. It is discretized with linear tetrahedra (about $58$k elements, $17.5$k nodes) and deforms by finite-strain J2 elastoplasticity with multilinear hardening (316L stainless steel, $E = 200$ GPa, $\sigma_y = 264$ MPa). The balloon is a folded Saint Venant--Kirchhoff membrane of nylon ($E = 850$ MPa) inflated by a follower pressure $\beta$ that expands the stent through the balloon--stent penalty interface, and the stent loads the wall through the strut--lumen interface of Appendix~\ref{app:formulation-contact}. The catheter and guidewire are not modelled: they position and stiffen the assembly rather than drive expansion.

\section{Derivation of the stress and consistent tangent}
\label{app:derivations}

The second Piola--Kirchhoff stress $\mathbf{S} = 2\,\partial\Psi/\partial\mathbf{C}$ inherits the split of \eqref{eq:energy} as $\mathbf{S} = \tfrac{1}{D}\left(J^{2}-1\right)\mathbf{C}^{-1} + J^{-2/3}\,\mathbb{P}:\bar{\mathbf{S}}$, with $\mathbb{P} = \mathbb{I} - \tfrac{1}{3}\mathbf{C}^{-1}\otimes\mathbf{C}$ the referential deviatoric projector, and the Cauchy stress follows as $\boldsymbol{\sigma} = J^{-1}\mathbf{F}\mathbf{S}\mathbf{F}^{\top}$. The fictitious stress is
\begin{equation}
\bar{\mathbf{S}} = c\,\mathbf{I} + 2k_1\sum_{i=1}^{2}\left\langle\bar{E}_i\right\rangle\exp\!\left(k_2\left\langle\bar{E}_i\right\rangle^{2}\right)\mathbf{H}_i,
\label{eq:stress}
\end{equation}
and differentiating once more, $\bar{\mathbb{C}} = 2\,\partial\bar{\mathbf{S}}/\partial\bar{\mathbf{C}}$, yields the fiber tangent \eqref{eq:tangent}.

The fiber tangent is
\begin{equation}
\bar{\mathbb{C}}_{\mathrm{f}}=4k_1\sum_{i=1}^{2}\mathcal{H}(\bar E_i)
(1+2k_2\bar E_i^2)\exp(k_2\bar E_i^2)\mathbf{H}_i\otimes\mathbf{H}_i,
\label{eq:tangent}
\end{equation}
where $\mathcal{H}$ is the Heaviside step. Each family's stress is continuous at recruitment, but its tangent jumps by $4k_1\mathbf{H}_i\otimes\mathbf{H}_i$. The assembled displacement Jacobian also includes geometric, volumetric, contact, and follower-load contributions.

\section{Adjoint derivation and computational cost}
\label{app:adjoint}

After history updates are substituted into the residual, an earlier displacement can affect later steps. Our~\eqref{eq:adjoint} includes these dependencies. The endpoint calculation instead holds recorded history and contact pairs fixed. Differentiating plastic history does not differentiate the discrete contact-pair search. The forward pass never assembles $\mathbf{K}_n$. Instead the transposed operator is applied as one reverse sweep through a single recorded assembly of the static residual at the recorded state, and recovered exactly as a sparse matrix by graph-colored vector--Jacobian probing (about $500$ colors, $1.5$k products at $1.4\times10^{5}$ degrees of freedom), which is factorized once per load step in double precision and shared across every outcome functional. Both functionals of one load step then cost about $30$ s against a forward deployment of $12$ min to $1$ h, so the adjoint cost is a small fixed multiple of the forward cost. Finite differences require a number of perturbed forward solves proportional to $\dim\boldsymbol{\theta}$, and each solve can evaluate all outcomes. Differentiating the unrolled explicit trajectory instead stores every step, so its memory footprint scales with the relaxation step count, which rises exactly where the problem is stiffest.

\section{Supplementary gradient analysis}
\label{app:pathologies}

\subsection{Finite-difference and complex-step reference accuracy}
\label{app:pathologies-fd}

The sensitivity of the fictitious stress to the stiffening coefficient is
\begin{equation}
\frac{\partial\bar{\mathbf{S}}}{\partial k_2}
= 2k_1\sum_{i=1}^{2}\left\langle\bar{E}_i\right\rangle^{3}
\exp\!\left(k_2\left\langle\bar{E}_i\right\rangle^{2}\right)\mathbf{H}_i .
\label{eq:dSdk2}
\end{equation}
Exponential stiffening can also degrade the finite-difference reference: a central difference has error $\mathcal{O}\!\left(h^2\right) + \mathcal{O}\!\left(\varepsilon_{\mathrm{mach}}\exp(k_2\bar{E}^2)/h\right)$. The complex-step derivative $\Im\!\left[f(\theta + i h)\right]/h$ avoids subtractive cancellation but requires an analytic function. We therefore use it to check the constitutive law on smooth branches away from recruitment transitions and any active numerical clamp. It does not supply a reference across those switches or define an additional backward-pass treatment.

\subsection{Reachability of the compressive fiber branch}

\label{app:pathologies-reachability}

Because $\bar{I}_1 \geq 3$ for any isochoric deformation, at $\kappa = 1/3$ the fiber strain \eqref{eq:fiberstrain} reduces to $\bar{E}_i = \tfrac{1}{3}(\bar{I}_1 - 3) \geq 0$ and the compressive branch of the Macaulay bracket is unreachable. Decreasing $\kappa$ permits compressive fiber strains for suitable deformations, but the number of points near recruitment also depends on loading and fiber orientation.

\subsection{Conditioning and volumetric locking}
\label{app:pathologies-locking}

The condition-number growth $\operatorname{cond}(\mathbf{K}) = \mathcal{O}(K/c)$ sits on top of the usual mesh dependence. Low-order tetrahedra incur volumetric locking: on the benchmark of Appendix~\ref{app:forward-validation}, tetrahedra and trilinear hexahedra agree to three digits in the plaque-free vessel and differ by up to $11\%$ in deployed lumen area once a near-incompressible plaque is present, entirely through locking in the plaque, which is why the tissue is discretized with hexahedra. Mixed or stabilized formulations can address locking, but are not used as backward-pass treatments here. Double-precision factorization improves the linear solve without correcting forward discretization error. The relative adjoint error is bounded by the condition number times the relative linear-solve residual, assuming an exact operator and right-hand side.

\section{Backward-pass treatment specifications}
\label{app:treatments}

We implement two backward-pass treatments and evaluate both in Section~\ref{sec:exp-ablations}: a chained adjoint that differentiates the plastic history across the recorded load steps, and a contact activation correction that supplies surrogate stiffness near contact activation. Neither changes the forward pass as the simulator validated in Appendix~\ref{app:forward-validation} is used for forward simulations. The chained adjoint extends the derivative, whereas the contact correction replaces part of it with a surrogate. All other reported gradients use the endpoint adjoint of Appendix~\ref{app:adjoint} without either treatment.

\paragraph{Chained plastic-history adjoint.} Each element of the stent and of the yielding plaque carries the history $\mathbf{h}=(\mathbf{F}_p^{-1},\alpha)$, the inverse plastic deformation gradient and the accumulated equivalent plastic strain. At recorded load step $n$, a single return map starts from the history $\mathbf{h}_{n-1}$ recorded at the previous step and is driven by the displacement $\mathbf{u}_n$. It gives both the stress in the residual and the updated history:
\begin{equation}
\mathbf{R}_n(\mathbf{u}_n,\mathbf{h}_{n-1};\boldsymbol{\theta},\beta_n)=\mathbf{f}_n,
\qquad
\mathbf{h}_n=\boldsymbol{\Phi}_n(\mathbf{u}_n,\mathbf{h}_{n-1};\boldsymbol{\theta}),
\qquad n=1,\ldots,N .
\label{eq:chain-constraints}
\end{equation}
Here $\mathbf{f}_n$ is the recorded residual force, held fixed as in the endpoint adjoint. Between recorded steps the forward pass applies the return map at every relaxation step. $\boldsymbol{\Phi}_n$ instead treats the whole increment from step $n-1$ to step $n$ as one return-map step, as an implicit load-stepping solver would. The chain uses $\boldsymbol{\Phi}_n$ only for its derivatives at the recorded states, and the approximation improves as more steps are recorded. On T2 the chain links the steps recorded at $t=0.5$, $1.0$, $1.5$, $1.8$, and $3.0$~s, and the history carried into the first of them is held fixed. For a functional $\mathcal{J}(\mathbf{u}_N;\boldsymbol{\theta})$ of the last recorded state, we adjoin the equilibria and updates with multipliers $\boldsymbol{\lambda}_n$ and $\boldsymbol{\xi}_n$ respectively. Setting the derivatives of the resulting Lagrangian with respect to the states to zero gives the backward sweep
\begin{align}
\mathbf{K}_N^{\top}\boldsymbol{\lambda}_N
&=\left(\frac{\partial\mathcal{J}}{\partial\mathbf{u}_N}\right)^{\!\top},
\qquad
\boldsymbol{\xi}_N=\mathbf{0},
\nonumber\\
\boldsymbol{\xi}_n
&=\left(\frac{\partial\boldsymbol{\Phi}_{n+1}}{\partial\mathbf{h}_n}\right)^{\!\top}\boldsymbol{\xi}_{n+1}
-\left(\frac{\partial\mathbf{R}_{n+1}}{\partial\mathbf{h}_n}\right)^{\!\top}\boldsymbol{\lambda}_{n+1},
\qquad n=N-1,\ldots,1,
\nonumber\\
\mathbf{K}_n^{\top}\boldsymbol{\lambda}_n
&=\left(\frac{\partial\boldsymbol{\Phi}_n}{\partial\mathbf{u}_n}\right)^{\!\top}\boldsymbol{\xi}_n,
\qquad n=N-1,\ldots,1,
\nonumber\\
\frac{\mathrm{d}\mathcal{J}}{\mathrm{d}\boldsymbol{\theta}}
&=\frac{\partial\mathcal{J}}{\partial\boldsymbol{\theta}}
-\sum_{n=1}^{N}\boldsymbol{\lambda}_n^{\top}\frac{\partial\mathbf{R}_n}{\partial\boldsymbol{\theta}}
+\sum_{n=1}^{N-1}\boldsymbol{\xi}_n^{\top}\frac{\partial\boldsymbol{\Phi}_n}{\partial\boldsymbol{\theta}} .
\label{eq:chain-sweep}
\end{align}
Eliminating $\boldsymbol{\xi}_n$ recovers the cross-step terms of \eqref{eq:adjoint} with the updates substituted into the residuals. The endpoint adjoint keeps only $\boldsymbol{\lambda}_N$ and the $n=N$ term of the gradient. Every term in \eqref{eq:chain-sweep} is a vector--Jacobian product, and no Jacobian of $\boldsymbol{\Phi}_n$ is formed. A reverse sweep through the recorded residual supplies the products with $\mathbf{K}_n^{\top}$ (probed and factorized once per step, Appendix~\ref{app:adjoint}), $\partial\mathbf{R}_n/\partial\boldsymbol{\theta}$, and $\partial\mathbf{R}_n/\partial\mathbf{h}_{n-1}$. A reverse sweep through one recorded application of the plastic update supplies the three products with $\boldsymbol{\Phi}_n$. Relative to the endpoint adjoint, the chain adds one transposed solve per earlier recorded step and functional, reusing that step's factorization. The return map is differentiated on the branch it takes at the recorded state, so the set of yielding elements is held fixed, the plastic analogue of the frozen contact set. The contact pairs are held fixed as well. We check each vector–Jacobian product separately against central finite differences, either on a single yielded element or at a single recorded load step (no chaining across steps). Over 62 comparisons the finite-difference-to-adjoint ratio lies between 0.89 and 1.08.

\paragraph{Contact activation correction.} The normal traction of \eqref{eq:contact}, $t_N=-\epsilon_N\langle -g_N\rangle$, has stiffness $\partial t_N/\partial g_N=\epsilon_N\mathcal{H}(-g_N)$, so a pair does not contribute to $\mathbf{K}_n$ until its gap closes. When the residual is recorded for the adjoint, both interfaces (balloon--stent and strut--lumen) keep the exact bracket in the residual value but differentiate it with the straight-through substitution
\begin{equation}
\epsilon_N\,\mathcal{H}(-g_N)\;\longrightarrow\;\epsilon_N\,\sigma(-g_N/w),
\qquad
\sigma(s)=\left(1+e^{-s}\right)^{-1},
\label{eq:contact-ste}
\end{equation}
and the candidate-pair search is widened from $g_N<0$ to $g_N<4w$, where the surrogate stiffness has decayed to $\sigma(-4)\,\epsilon_N\approx0.018\,\epsilon_N$. Pairs in the widened band carry no force but contribute surrogate stiffness. Active pairs keep their force, with surrogate stiffness $\epsilon_N/2$ at $g_N=0$, and approach the exact value once the penetration exceeds a few $w$. The residual value, and therefore the recorded state and $\mathbf{f}_n$, are unchanged. Because the contact force does not depend on $\boldsymbol{\theta}$ directly, the substitution enters the adjoint only through $\mathbf{K}_n^{\top}$. Friction does not enter, since the tangential traction vanishes from the residual at zero velocity. Setting $w=0$ recovers the frozen-set derivative, and Section~\ref{sec:exp-ablations} presents results for $w$ from $0.1$ to $5$~$\mu$m. The correction smooths the stiffness of the frozen active set \eqref{eq:active-set} but does not differentiate the pair search. It is therefore not an exact active-set derivative, and pairs whose gap exceeds $4w$ at the recorded state remain invisible to the adjoint.

\section{Solver tolerances, mesh convergence, and hyperparameters}
\label{app:hyperparams}

\paragraph{Testbeds and repeat protocol.} The T1 cube comprises $2\times2\times2$ trilinear hexahedra under symmetric uniaxial stretch. The tetrahedral variant splits each hexahedron into six tetrahedra. The compressible HGO law admits a closed-form solution for the cube's equilibrium and nominal stress, while the single-layer tube is loaded by lumen pressure. T2 uses a straight $3$ mm vessel with prescribed layerwise fiber angles, a closed-cell stent, and a folded balloon; its stenosed configuration contains yielding plaque. We record deployment at fixed load steps as pressure ramps to its calibrated peak. Finite-difference comparisons use six to nine identical deployments at T2 study points and four at $\kappa=0$ and on T3, including the reference. The study comprise $960$ T2 and $57$ T3 deployments. We report forward and adjoint solve counts alongside wall time.

\paragraph{Forward relaxation.} All solids share one mm--kg--s unit system with stresses in kPa and are advanced by a mass-scaled, semi-implicit Euler integrator in single precision. The stent's $0.02$--$0.03$ mm tetrahedra set the stable step, $\Delta t = 4\times10^{-6}$ s, at a mass scaling of $10^{7}$ for the benchmark deployments of Appendix~\ref{app:forward-validation} ($1.5\times10^{6}$ steps for a 3 s inflation and 3 s deflation, kinetic-to-internal energy ratio below $5\%$). Deployments recorded for the adjoint on T2 use $\Delta t = 10^{-5}$ s, a mass scaling of $4\times10^{7}$ on the stent, $10^{6}$ on the balloon, and $10^{5}$ on the wall (the wall's own stability limit with a tenfold margin), so that the wall relaxes within a load step. The residual dynamic-force fraction is recorded per region at every load step and is $1$--$2\%$ on the plaque-free wall and $5$--$9\%$ on the wall and $20\%$ in the plaque at the stenosed peak, where an additional held-load relaxation is applied before the state is used.

The residual dynamic force is $\mathbf{R}=-\mathbf{M}(\mathbf{a}+\eta\mathbf{v})$, where $\mathbf{M}$ is the mass matrix, $\mathbf{a}$ and $\mathbf{v}$ are acceleration and velocity, and $\eta$ denotes damping, distinct from material stiffness $c$.

\paragraph{Adjoint solve.} The static residual is recorded once per load step in single precision with velocities zeroed and the contact pairs frozen. The exact transposed tangent is recovered by graph-colored probing (Appendix~\ref{app:adjoint}), verified by re-applying it to a random vector (relative error below $10^{-6}$), factorized in double precision, and used for the direct adjoint solve reported in Section~\ref{sec:exp-correctness}..

\paragraph{Discretizations.} The benchmark tissue is a structured mesh of $23{,}040$ trilinear hexahedra ($29{,}040$ nodes) with minimum cell thickness of $0.074$ mm. The stent has about $58$k linear tetrahedra. The T2 sweep study uses a coarser tissue mesh ($32$ rays by $90$ axial stations, $11{,}520$ hexahedra) chosen from a measured element aspect-ratio table. Re-running the stenosed reference on the finer $48\times120$ mesh (twice the element count, six repeats, finite-difference pairs on the six parameters of the contact-side sweeps) keeps the six largest sensitivities of each layer's mean stress at the peak in the same order (e.g., media stress: media $c$ $+83 \to +83\%$, intima $\kappa$ $-26 \to -21\%$, media $\kappa$ $-20 \to -19\%$; intima stress: intima $\kappa$ $-54 \to -52\%$, intima $k_1$ $+43 \to +42\%$, intima $D$ $-26 \to -26\%$; adventitia stress: adventitia $\kappa$ $-176 \to -186\%$). Below them, the plaque $G$ on the adventitia stress halves ($-6.9 \to -3.7\%$) but it doesn't change the ordering of the sensitivities. The $48\times120$ finite-difference ratios are the last row of Table~\ref{tab:t2-contact}. 

\section{Additional gradient-validation results}
\label{app:gradient-validation}

We evaluate the accuracy of our adjoint implementation against central finite differences for T2 and T3 (material parameters perturbed by $\pm5\%$) against a double-precision derivative of the closed-form solution for the T1 cube, and against central finite differences for the T1 tube.

The functionals we differentiate are lumen area at the stenosis throat and volume-mean von Mises stress in each wall layer.

We sweep parameters to characterize the regimes in which the adjoint method fails. The stretch sweep traverses exponential and lock-up regimes, $k_2$ changes the strength of exponential stiffening, dispersion $\kappa \in [0,1/3]$ and fiber angle move fibers toward, across, or away from the tension-only recruitment surface, probing the pathology in which arterial fibers switch on at a threshold strain. Finally, the bulk-to-shear sweep approaches the incompressible limit and tests the conditioning predicted by pathology (ii).

On T2, we sweep contact stiffness, friction, and the plaque-to-media modulus ratio. These sweeps aim to test the validity of the adjoint's frozen contact set assumption.

Our method is faster and more accurate compared to existing methods for calculating gradients (Table~\ref{tab:t1-methods}). Reverse-mode autodiff and the adjoint agree with the closed form to the same single-precision level ($2.5\times10^{-5}$ median, $10^{-3}$ worst), as the identical-truncation check of Section~\ref{sec:exp-adjoint} predicts. The $\pm5\%$ forward difference is $20\times$ less accurate in the median, and its worst entries ($0.3$--$2.6$) are the parameters whose sensitivity is small ($k_2$ at $\kappa = 1/3$, $10^{-9}$ in the lateral stretch), where the difference of two relaxed forwards is dominated by their rounding. The adjoint's gradient costs $0.4$ s on top of the relaxation, against $271$ s for the unrolled differentiation and nine further relaxations for the finite difference.
\begin{table}[h]
\centering
\footnotesize
\caption{Gradient accuracy and cost on the constitutive cube (T1, $2\times2\times2$ hexahedra). Each method is measured against the double-precision derivative of the closed-form solution over five material parameters and three outcomes and $23$ sweep points. Time includes the $N = 200\,000$-step relaxation ($98$ s)}
\label{tab:t1-methods}
\setlength{\tabcolsep}{3.5pt}
\begin{tabular}{@{}lccccc@{}}
\toprule
gradient & median error & worst error & forwards & states held & time \\
\midrule
closed form, fp64 (reference) & $3.8\times10^{-9}$ & $3.6\times10^{-5}$ & $0$ & -- & -- \\
finite difference, $\pm5\%$ & $5.6\times10^{-4}$ & $2.6$ & $2\dim\boldsymbol{\theta} = 10$ & $1$ & $979$ s \\
unrolled autodiff & $2.5\times10^{-5}$ & $1.9\times10^{-3}$ & $1$ & $N$ ($1.5$ GiB) & $369$ s \\
equilibrium adjoint (ours) & $2.5\times10^{-5}$ & $9.9\times10^{-4}$ & $1$ & $1 + \mathbf{K}$ & $98 + 0.4$ s \\
\bottomrule
\end{tabular}
\end{table}

\paragraph{Results on T1.} Tables~\ref{tab:t1-cube} and \ref{tab:t1-tube} give the accuracy map on the constitutive testbeds. The adjoint largely matches the closed-form derivative on the tetrahedral cube and the finite difference on the tube. We maintain a gradient cosine similarity above 0.99999 everywhere. The gradient remains robust in the exponential and lock-up regimes ($\lambda = 1.4$), and when the fibers are placed in compression. The two peak-stress cases are outliers in which the direct and adjoint term of \eqref{eq:adjoint} cancel to $1\%$  of their size, so a small error in either part compounds into a larger percentage of the remainder.

\begin{table}[h]
\centering
\small
\caption{T1 constitutive cube: worst relative gradient error against closed-form derivative over the five material parameters ($c$, $k_1$, $k_2$, $\kappa$, $D$) and three outcomes (reaction force, two lateral stretches)}
\label{tab:t1-cube}
\setlength{\tabcolsep}{4pt}
\begin{tabular}{@{}llll@{}}
\toprule
swept parameter & range & hexahedra & tetrahedra\\
\midrule
stretch $\lambda$ & $1.02$--$1.4$ & $3.5\times10^{-4}$ & $2.8\times10^{-4}$\\
fiber angle $\gamma$ & $0$--$90^\circ$ & $2.1\times10^{-4}$ & $5.4\times10^{-5}$\\
dispersion $\kappa$ & $0$--$1/3$ & $9.8\times10^{-4}$ & $2.1\times10^{-4}$\\
stiffening $k_2$ & $20$--$400$ & $2.1\times10^{-4}$ & $6.5\times10^{-5}$\\
compressibility $D$ & $10^{-1}$--$10^{-4}$ & $3.2\times10^{-4}$ & $6.5\times10^{-4}$\\
compressibility $D$ & $10^{-5}$ (relaxation-limited) & $9.2\times10^{-3}$ & $5.8\times10^{-3}$\\
\bottomrule
\end{tabular}
\end{table}

\begin{table}[h]
\centering
\small
\caption{T1 single-layer tube under lumen pressure: relative gradient error of the adjoint against central finite differences of the full relaxed forward, for the wall's $c$ and $k_1$, by outcome.}
\label{tab:t1-tube}
\setlength{\tabcolsep}{4pt}
\begin{tabular}{@{}lllp{5.2cm}@{}}
\toprule
sweep & values & lumen area, displacement & peak stress \\
\midrule
dispersion $\kappa$ & $0$, $0.1$, $0.2$, $1/3$ & $\leq 0.11\%$ & $\leq 0.75\%$ \\
pressure & $5$, $20$, $40$ kPa & $\leq 0.12\%$ & $\leq 0.1\%$, except $4.3\%$ at $5$ kPa ($c$) \\
stiffening $k_2$ & $20$, $200$, $400$ & $\leq 0.13\%$ & $\leq 0.3\%$, except $3.5\%$ at $k_2 = 20$ ($k_1$) \\
\bottomrule
\end{tabular}
\end{table}

\paragraph{} The precision of the FD is itself limited by the run-to-run noise of the forward model. Single-precision GPU reductions with nondeterministic summation order seed differences at the level of round-off, and two features of the contact model amplify them: the regularized Coulomb friction can switch a strut between sticking and sliding on the lumen, and the discrete contact footprint can gain or lose a collocation point (\eqref{eq:active-set}). Thus, two deployments with identical inputs can differ in lumen gain or wall stress.

We therefore run the reference T2 deployments six to nine times and define the floor as the standard deviation of the outcome (lumen gain or a layer's mean stress). We call a pair \emph{resolvable} when its observed or adjoint-predicted change exceeds three floors, and we quote the signal-to-noise ratio (observed effect over floor) with every comparison.

The lumen gain is insensitive to wall parameters: the nearly inextensible $3$ mm balloon prescribes the stent diameter. Wall stiffness determines the contact force rather than the lumen area, so we study layer-mean stresses as outcomes.

\paragraph{Results on T2.} Table~\ref{tab:t2-reference} gives, at the reference point, the sensitivity of each layer's mean stress to that layer's wall parameters and the finite-difference-to-adjoint ratio of each; Table~\ref{tab:t2-points} shows the material sweeps against the media stress.

The adjoint consistently misses the intima, the layer in contact with the struts. Against the media stress, its dispersion, a pure adjoint-term parameter, is over-predicted by $16$--$33\%$ at six stenosed points (ratios $0.67$--$0.84$, signal-to-noise $9$--$16$) and by $59\%$ on the plaque-free $2.5$ mm vessel, sits within $10\%$ at four points ($\mu = 0$, $\epsilon_N = 3\times10^{4}$, modulus ratio $15$, and the $48\times120$ mesh at the peak), and is under-predicted at modulus ratio $152$. Against the intima's own stress it is under-predicted by $30\%$ at $1.0$ s (signal-to-noise $135$) and exact at the peak. Its $k_1$ is under-predicted by $25$--$30\%$ (ratios $1.32$--$1.37$). Halving the step, removing friction, and the plastic-history treatment of Section~\ref{sec:exp-ablations} fail to rescue the miss.

The media $k_2$ is noisy in the regime where its direct and adjoint terms cancel by $80$--$95\%$, so a percent-level error in either becomes a large error in the remainder. And the plaque modulus ($c$ and $c_{01}$ of the yielding plaque scaled together) gives no consistent verdict ($0.5$--$6.4$ across the loaded steps of the material-sweep points, $17$ at modulus ratio $152$ after deflation). Further, its sensitivity is path-dependent through the plaque's yield and the strut--plaque contact, which we add in Section~\ref{sec:exp-ablations}. At $\kappa = 1/3$ the intima $\kappa$ ratio flips sign because the finite difference straddles the isotropic limit, where the switch does not exist.

\begin{table}[h]
\centering
\footnotesize
\caption{T2 reference point ($60\%$ eccentric stenosis, $\kappa = 0.2$): each layer's mean von Mises stress against its material parameters. Sensitivity is the adjoint's $\theta\,\partial J/\partial\theta$ at the peak pressure. FD/AD is the central finite difference over the adjoint at the load step $t$ at which the parameter is best resolved, with the signal-to-noise ratio of that pair (observed effect over the repeat floor). Pairs use $\pm5\%$ steps ($\pm0.02$ for $\kappa$); sensitivities marked $^\dagger$ are unresolved at $\pm5\%$ at every step (below the floor, or for the media $D$ a rejected one-sided pair) and use $\pm20\%$.}
\label{tab:t2-reference}
\setlength{\tabcolsep}{3pt}
\begin{tabular}{@{}l rrrr rrrr rrrr@{}}
\toprule
& \multicolumn{4}{c}{\textsc{intima stress}} & \multicolumn{4}{c}{\textsc{media stress}} & \multicolumn{4}{c}{\textsc{adventitia stress}} \\
\cmidrule(lr){2-5} \cmidrule(lr){6-9} \cmidrule(lr){10-13}
leaf & sens.\ \% & FD/AD & SNR & $t$ (s) & sens.\ \% & FD/AD & SNR & $t$ (s) & sens.\ \% & FD/AD & SNR & $t$ (s) \\
\midrule
$c$ & $+8$ & 0.84 & 6 & 1.0 & $+83$ & 1.00 & 61 & 1.0 & $+15$ & 0.92 & 27 & 1.0 \\
$k_1$ & $+43$ & 1.37 & 30 & 1.0 & $+8$ & 0.93$^\dagger$ & 11 & 1.8 & $+38$ & 0.96 & 25 & 1.0 \\
$k_2$ & $+5$ & 1.31$^\dagger$ & 6 & 1.8 & $+7$ & 0.73$^\dagger$ & 6 & 1.8 & $+58$ & 1.02 & 9 & 1.0 \\
$\kappa$ & $-54$ & 1.44 & 135 & 1.0 & $-20$ & 0.86 & 14 & 1.5 & $-176$ & 0.99 & 160 & 1.0 \\
$D$ & $-26$ & 1.10 & 8 & 1.0 & $-0$ & --$^\dagger$ & 2 & 1.0 & $-34$ & 1.02 & 5 & 1.8 \\
\bottomrule
\end{tabular}
\end{table}

\begin{table}[h]
\centering
\footnotesize
\caption{T2 material sweeps: finite difference over adjoint for the mean media stress at the peak pressure, with the signal-to-noise ratio in brackets. The cosine is over the resolvable sensitivities. ``--'' marks a parameter whose effect is below three floors at this step, $^\ast$ a pair rejected although above the floor (one endpoint one-sided or off the main unfolding path), n/a a parameter the vessel does not have; $^\ddagger$ marks the pair that straddles the isotropic limit $\kappa = 1/3$, where the recruitment switch ceases to exist. Repeats count the identical deployments on the main path, reference included.}
\label{tab:t2-points}
\setlength{\tabcolsep}{2.5pt}
\begin{tabular}{@{}l c c c c c c c c@{}}
\toprule
point & repeats & floor (\%) & media $c$ & media $\kappa$ & intima $k_1$ & intima $\kappa$ & plaque $G$ & cosine \\
\midrule
$\kappa = 0$ (benchmark) & 4 & 0.83 & 1.03 [8] & 0.87 [9] & -- & -- & -- & 0.97 \\
$\kappa = 0.1$ & 7 & 0.13 & 0.89 [51] & 1.11 [55] & 0.87 [3] & --$^\ast$ & 5.55 [7] & 0.96 \\
$\kappa = 0.2$ (reference) & 8 & 0.25 & 1.02 [34] & 0.86 [14] & 1.32 [4] & 0.76 [16] & 2.81 [6] & 0.99 \\
$\kappa = 1/3$ & 6 & 0.30 & 1.01 [31] & 0.97 [5] & -- & $-2.41$ [7]$^\ddagger$ & -- & 0.97 \\
media $k_2 = 20$ & 7 & 2.1 & 0.95 [4] & -- & -- & -- & -- & -- \\
media $k_2 = 200$ & 7 & 0.59 & 1.21 [12] & 0.96 [26] & -- & -- & -- & 0.99 \\
media $D \times 10$ & 7 & 0.37 & 1.05 [22] & 1.02 [7] & -- & 0.80 [15] & 1.49 [4] & 0.99 \\
media $D \div 10$ & 6 & 0.25 & 1.00 [31] & 1.07 [22] & -- & 0.72 [12] & 3.44 [8] & 0.98 \\
media fibers $45^\circ$ & 7 & 0.42 & 1.00 [20] & --$^\ast$ & -- & 0.75 [9] & -- & 1.00 \\
media fibers $60^\circ$ & 6 & 0.53 & 1.09 [17] & 0.91 [5] & -- & 0.84 [9] & -- & 1.00 \\
plaque-free, $3$ mm & 7 & 2.1 & --$^\ast$ & 13.4 [5] & 7.31 [6] & -- & n/a & 0.50 \\
plaque-free, $2.5$ mm & 7 & 0.92 & 1.08 [10] & -- & -- & 0.41 [5] & n/a & 0.94 \\
\bottomrule
\end{tabular}
\end{table}

The mean stress of each layer is governed by few parameters (relative sensitivity $|\theta\,\partial J/\partial\theta|/J \geq 5\%$ at the reference point). No wall parameter governs the lumen area above its floor.

\paragraph{Contact-side sweeps.} Table~\ref{tab:t2-contact} follows the intima $\kappa$ (the systematic miss) and the media $c$ (the control) across the three contact sweeps. Adding friction reduces accuracy: the frozen active set is unable to resolve the friction-held configuration. The penalty stiffness moves the media $c$ by the conditioning effect of a penalty $10^{4}$ times stiffer than the wall it presses on. As predicted by the contact mechanism pathology, stiffness contrast at the calcific interface reduces the accuracy of intima $\kappa$. At modulus ratio $152$ plaque plasticity is excluded: the plaque's yield stress scales with its modulus, and no plaque element yields in any deployment at that ratio. The mechanism is not identified: there the $1$~$\mu$m activation correction of Section~\ref{sec:exp-ablations} moves the peak-step FD/adjoint from $1.91$ to $1.19$ but leaves the $1.0$ s and $3.0$ s misses ($4.5$ and $2.0$) in place. Refining the tissue mesh to $48\times120$ (twice the element count, six repeats; last row) leaves the media $c$ and $\kappa$ at $0.95$ and $0.93$ and moves the intima $\kappa$ from $0.76$ to $0.95$ (SNR $=34$). On the finer mesh the systematic miss is confined to the mid-ramp step ($0.67\to0.72$ at $1.0$ s, where the media $c$ control also falls, $1.00\to0.87$), so the discretization sets much of its size. The plaque $G$ pair is not resolved on the finer mesh, whose balloon-path check was calibrated on the coarse mesh.

\begin{table}[h]
\centering
\footnotesize
\caption{T2 contact-side sweeps at $\kappa = 0.2$: finite difference over adjoint (FD/AD) for the mean media stress at the peak pressure, with the SNR of each pair. The modulus ratio is the plaque-to-media shear-modulus ratio $G_{\mathrm{plaque}}/G_{\mathrm{media}}$ with the plaque's yield stress scaled with its modulus; The reference is $\mu = 0.2$, $\epsilon_N = 10^4$ kPa/mm and modulus ratio $1.5$. Where the observed effect is below three floors the ratio is omitted and only the SNR is listed; $^\ast$ marks a pair that is not a derivative because one endpoint left the main unfolding path or the pair was one-sided.}
\label{tab:t2-contact}
\setlength{\tabcolsep}{1.7pt}
\begin{tabular}{@{}l c c cc cc cc cc cc c@{}}
\toprule
& & & \multicolumn{2}{c}{media $c$} & \multicolumn{2}{c}{media $\kappa$} & \multicolumn{2}{c}{intima $k_1$} & \multicolumn{2}{c}{intima $\kappa$} & \multicolumn{2}{c}{plaque $G$} & \\
\cmidrule(lr){4-5} \cmidrule(lr){6-7} \cmidrule(lr){8-9} \cmidrule(lr){10-11} \cmidrule(lr){12-13}
point & repeats & floor (\%) & FD/AD & SNR & FD/AD & SNR & FD/AD & SNR & FD/AD & SNR & FD/AD & SNR & cosine \\
\midrule
reference & 8 & 0.25 & 1.02 & 34 & 0.86 & 14 & 1.32 & 4 & 0.76 & 16 & 2.81 & 6 & 0.99 \\
friction $\mu = 0$ & 9 & 0.63 & 0.81 & 11 & 0.48 & 3 & -- & 0.5 & 0.90 & 7 & -- & 0.2$^\ast$ & 1.00 \\
friction $\mu = 0.5$ & 7 & 1.8 & 0.96 & 4 & -- & 2.4 & -- & 0.2 & -- & 2.0 & -- & 2.1$^\ast$ & -- \\
$\epsilon_N = 3\times10^4$ & 6 & 0.56 & 0.88 & 13 & 0.74 & 5 & -- & 0.1 & 1.00 & 8 & -- & 2.0$^\ast$ & 1.00 \\
$\epsilon_N = 10^5$ & 6 & 0.24 & 1.03 & 35 & 0.87 & 14 & -- & 1.5 & 0.67 & 16 & -- & 3.9$^\ast$ & 0.99 \\
ratio $15$ & 7 & 0.47 & 0.98 & 17 & -- & 0.2 & -- & 3.6$^\ast$ & 1.00 & 5 & -- & 4.4$^\ast$ & 1.00 \\
ratio $152$ & 7 & 0.10 & 1.00 & 62 & 0.89 & 7 & 1.66 & 5 & 1.91 & 31 & 0.74 & 45 & 0.98 \\
ratio $1521$ & 7 & 3.2 & -- & 1.5 & -- & 0.3 & -- & 2.8 & -- & 1.1 & -- & 0.4 & -- \\
mesh $48\times120$ & 6 & 0.12 & 0.95 & 67 & 0.93 & 31 & 0.66 & 4 & 0.95 & 34 & -- & 14$^\ast$ & 1.00 \\
\bottomrule
\end{tabular}
\end{table}

\paragraph{Results on T3.} On the patient-specific artery, whose material sweeps do not exist, all $15$ wall parameters were checked against $\pm5\%$ pairs with four identical deployments (Table~\ref{tab:t3-verify}). Of the nine layer stress to layer parameter comparisons that resolve at the peak, the adventitia $\kappa$ agrees to $1\%$ at signal-to-noise $46$ and its $k_2$ to $3\%$, the media $\kappa$ and $c$ to $12$--$13\%$, the intima $\kappa$, $k_1$ and $D$ to $6$--$13\%$ in their own layer. The misses are the media $k_1$ ($2.2$ at signal-to-noise $5$, the parameter at the edge of resolvability on T2) and the adventitia $k_1$ ($0.63$ at signal-to-noise $3$). The lumen area is resolvable here for the first time, because the calcified lesion is not dilated by the $3$ mm balloon at any pressure we applied (Table~\ref{tab:t3-pressure}) and the wall carries the load. The intima $\kappa$ moves the throat area at signal-to-noise $47$ at $1.0$ s and $10$ at the peak with a finite-difference-to-adjoint ratio of $2.6$ at both, the intima $k_1$ at $10$ and $4$ with ratios $1.7$ and $4.0$. With the $1$ $\mu$m activation correction of Section~\ref{sec:exp-ablations} the two peak ratios become $0.77$ and $1.26$, but the $1.0$ s intima $\kappa$ ratio (signal-to-noise $47$) only moves to $2.1$ and the peak ratio of the mean media stress to the intima $\kappa$ changes sign ($0.39\to-0.19$, signal-to-noise $7$), so a width tuned on T2 does not transfer to T3. It is the contact-layer miss of T2 on the outcome that matters clinically.

\begin{table}[h]
\centering
\footnotesize
\caption{T3 patient vessel at the peak pressure ($\kappa = 0.2$): each layer's mean von Mises stress against its own material parameters, sensitivity $\theta\,\partial J/\partial\theta / J$ (\%), FD/AD and SNR of the $\pm5\%$ pair (four identical deployments, reference included). Where the pair is below three floors only its SNR is listed; $^\ast$ marks a one-sided pair, which is not a derivative.}
\label{tab:t3-verify}
\setlength{\tabcolsep}{4pt}
\begin{tabular}{@{}l rrr rrr rrr@{}}
\toprule
& \multicolumn{3}{c}{\textsc{intima stress}} & \multicolumn{3}{c}{\textsc{media stress}} & \multicolumn{3}{c}{\textsc{adventitia stress}} \\
\cmidrule(lr){2-4} \cmidrule(lr){5-7} \cmidrule(lr){8-10}
leaf & sens.\ \% & FD/AD & SNR & sens.\ \% & FD/AD & SNR & sens.\ \% & FD/AD & SNR \\
\midrule
$c$ & $+7$ & -- & 0.9 & $+67$ & 1.13 & 12 & $+21$ & -- & 2.3 \\
$k_1$ & $+41$ & 0.87 & 5 & $+16$ & 2.21 & 5 & $+40$ & 0.63 & 3 \\
$k_2$ & $+6$ & -- & 1.9 & $+17$ & -- & 0.1 & $+48$ & 0.97 & 6 \\
$\kappa$ & $-56$ & 0.94 & 16 & $-69$ & 0.88 & 19 & $-198$ & 0.99 & 46 \\
$D$ & $-24$ & 1.08 & 4 & $-1$ & -- & 7.5$^\ast$ & $-21$ & -- & 2.2 \\
\bottomrule
\end{tabular}
\end{table}

\begin{table}[h]
\centering
\footnotesize
\caption{T3 pressure sweep (Conway materials, $\kappa = 0$, the $3$ mm balloon): throat lumen area at the peak and after deflation. The pre-procedure throat is $1.2$ mm$^2$; the follow-up OCT, after a $4.5$ mm clinical balloon, shows $5.5$ mm$^2$ over the stented segment. The device does not dilate the calcified lesion at any of these pressures.}
\label{tab:t3-pressure}
\setlength{\tabcolsep}{6pt}
\begin{tabular}{@{}lrrrr@{}}
\toprule
$p$ (kPa) & 857 & 1100 & 1400 & 1800 \\
\midrule
throat lumen at the peak (mm$^2$) & 1.51 & 1.75 & 2.10 & 2.78 \\
throat lumen, deflated (mm$^2$) & 1.48 & 1.69 & 1.96 & 2.44 \\
recoil (\%) & 1.6 & 3.5 & 6.5 & 12.2 \\
\bottomrule
\end{tabular}
\end{table}

\section{Sources of adjoint error}
\label{app:adjoint-error-sources}

The results illustrate the three mechanisms in Section~\ref{sec:theory-pathologies}, each of which can make equilibrium sensitivities differ from finite differences of the full deployment.

Nonsmooth switching occurs when parameter perturbations activate or release contact pairs that the endpoint adjoint holds fixed. This contributes to errors in intima sensitivities beneath the stent struts. Smoothing contact activation reduces some discrepancies, but the improvement depends on deployment conditions and smoothing width (Section~\ref{sec:exp-ablations}). At a plaque-to-media modulus ratio of $152$, intima-stress errors persist after correction, leaving part of the discrepancy unexplained. Fiber recruitment also changes under perturbation, but its influence on media sensitivities is limited here because the media's fiber stiffness is only about $1\%$ of its ground-substance stiffness.

State mismatch arises because the endpoint adjoint assumes equilibrium and freezes plastic history. The recorded stent retains residual forces of $20$--$40\%$ of its internal force under load and $60$--$80\%$ after deflation. Pressure sensitivities depend on the response of this nonequilibrated device: their signs are usually correct at the held peak, but their magnitudes are typically $2$--$3\times$ too small, with larger underestimation after deflation (Section~\ref{sec:exp-clinical}). Propagating plastic history improves some plaque-modulus sensitivities but does not correct the force imbalance or recover the stent's yield-stress sensitivity (Section~\ref{sec:exp-ablations}).

Gradient cancellation amplifies errors when the direct and adjoint contributions nearly offset one another. For media $k_2$, these terms cancel by $80$--$95\%$, so small errors in either contribution become large relative errors in the remaining sensitivity. This also explains the peak-stress outliers on T1.

Near-incompressibility does not emerge as the dominant error source in these tests: increasing the media bulk modulus tenfold leaves FD/adjoint ratios of $c$ and $\kappa$ at $1.00$--$1.11$. Accurate solution of the adjoint system, however, does not remove switching, state mismatch, or cancellation errors.

\section{Parameter-wise gradient errors}
\label{app:unrolled-validation}

Table~\ref{tab:unrolled-endpoints} compares peak-wall-stress sensitivities from the equilibrium adjoint and unrolled autodiff on the T1 tube after $32{,}000$ relaxation steps. Relative errors are computed against the adjoint at a relaxed reference state.

The equilibrium adjoint agrees more closely with the reference for every parameter and compressibility setting. At $D=10^{-2}$~kPa$^{-1}$, the compressibility sensitivity dominates both methods' errors. At $D=10^{-3}$~kPa$^{-1}$, the adjoint's mean error is approximately $24\times$ lower. The stiffest tube ($D=10^{-4}$~kPa$^{-1}$) remains incompletely relaxed, so its comparison also reflects forward-state convergence. These results establish closer agreement with an equilibrium reference. On the independent T1 cube benchmark, both methods achieve the same median error against closed-form derivatives (Table~\ref{tab:t1-methods}).

\begin{table}[t]
\centering
\small
\caption{Relative peak-stress gradient errors at $32{,}000$ steps against a relaxed adjoint reference. Bold marks lower errors.}
\label{tab:unrolled-endpoints}
\setlength{\tabcolsep}{4pt}
\begin{tabular}{@{}clrrrrr@{}}
\toprule
$D$ (kPa$^{-1}$) & Method & $c$ & $k_1$ & $k_2$ & $D$ & Mean \\
\midrule
\multirow{2}{*}{$10^{-2}$} & Autodiff & $2.1\!\times\!10^{-4}$ & $6.1\!\times\!10^{-4}$ & $1.0\!\times\!10^{-3}$ & $4.0\!\times\!10^{-1}$ & $1.0\!\times\!10^{-1}$ \\
 & Adjoint & $\bm{9.3\!\times\!10^{-6}}$ & $\bm{3.0\!\times\!10^{-5}}$ & $\bm{3.0\!\times\!10^{-5}}$ & $\bm{2.5\!\times\!10^{-3}}$ & $\bm{6.5\!\times\!10^{-4}}$ \\
\midrule
\multirow{2}{*}{$10^{-3}$} & Autodiff & $3.9\!\times\!10^{-4}$ & $1.5\!\times\!10^{-3}$ & $1.0\!\times\!10^{-3}$ & $6.1\!\times\!10^{-4}$ & $8.8\!\times\!10^{-4}$ \\
 & Adjoint & $\bm{2.5\!\times\!10^{-5}}$ & $\bm{5.8\!\times\!10^{-5}}$ & $\bm{4.2\!\times\!10^{-5}}$ & $\bm{1.9\!\times\!10^{-5}}$ & $\bm{3.6\!\times\!10^{-5}}$ \\
\midrule
\multirow{2}{*}{$10^{-4}$} & Autodiff & $3.3\!\times\!10^{-3}$ & $3.3\!\times\!10^{-2}$ & $1.3\!\times\!10^{-2}$ & $4.0\!\times\!10^{-3}$ & $1.3\!\times\!10^{-2}$ \\
 & Adjoint & $\bm{2.8\!\times\!10^{-6}}$ & $\bm{2.2\!\times\!10^{-5}}$ & $\bm{1.0\!\times\!10^{-5}}$ & $\bm{3.4\!\times\!10^{-6}}$ & $\bm{9.5\!\times\!10^{-6}}$ \\
\bottomrule
\end{tabular}
\end{table}

\section{Limitations and Future Work}
\label{app:limitationsandfuture}

\subsection{Limitations}
\label{sec:exp-limitations}
\textbf{Contact.} DiffStent enforces contact with a penalty: a node that enters another body's contact band is pushed back by a spring force proportional to its penetration. The spring stiffness must be tuned to the load, since a soft spring lets struts penetrate the wall and a stiff one shortens the stable time step, and its on--off activation is the nonsmooth switch that the endpoint adjoint freezes and the contact surrogate only partly corrects (Section~\ref{sec:exp-ablations}). A smooth barrier formulation such as incremental potential contact \citep{li2020incremental} would prevent penetration without a hand-tuned stiffness and make the contact force differentiable across activation.

\textbf{Inverse problems.} Our material identification and procedural optimization results more information than clinical imaging can provide. The identification procedure inputs the displacement of every T3 tissue node at peak pressure, with no noise beyond the scatter between identical deployments, whereas intravascular and angiographic imaging resolve only the lumen and part of the wall. Even so, the estimates remain $3$--$23\%$ off (Section~\ref{sec:exp-clinical}). The pressure optimization assumes that the wall's material parameters are known. Using either in patients requires extraction of patient-specific material properties using inverse finite element optimization techniques (Section~\ref{sec:future}).

\subsection{Future Work}
\label{sec:future}
We plan to extend DiffStent in two sequential directions. First, we aim to infer patient-specific lesion properties from paired pre- and post-intervention images in retrospective clinical cohorts using inverse FE optimization. These estimates would allow us to study how lesion mechanics vary across patients and relate to stent expansion. Second, we aim to use the inferred property maps as supervision for an inference model that predicts spatially varying mechanical properties directly from anatomical imaging. Such a model could reduce the need for iterative parameter fitting in new patients and improve patient outcomes by providing material estimates for simulating deployment before intervention.

\end{document}